\documentclass[aps, prstab, reprint, floatfix, showpacs]{revtex4-1}

\usepackage{graphicx}
\usepackage{upgreek}
\usepackage{amsmath}
\usepackage{amssymb}
\usepackage{multirow}
\usepackage{notoccite}
\usepackage{threeparttable}
\newcommand{\unit}[1]{\ensuremath{\, \mathrm{#1}}}
\providecommand{\e}[1]{\ensuremath{\times 10^{#1}}}

\usepackage{array}
\newcolumntype{x}[1]{>{\centering\arraybackslash\hspace{0pt}}p{#1}}

\begin{document}
\title{Nitrogen-Doped 9-Cell Cavity Performance in a Test Cryomodule for LCLS-II}
\thanks{Work supported by the US DOE and the LCLS-II Project}

\author{D. Gonnella}
\email{dg433@cornell.edu}
\affiliation{CLASSE, Cornell University, Ithaca, NY 14853, USA}
\author{R. Eichhorn}
\affiliation{CLASSE, Cornell University, Ithaca, NY 14853, USA}
\author{F. Furuta}
\affiliation{CLASSE, Cornell University, Ithaca, NY 14853, USA}
\author{M. Ge}
\affiliation{CLASSE, Cornell University, Ithaca, NY 14853, USA}
\author{D. Hall}
\affiliation{CLASSE, Cornell University, Ithaca, NY 14853, USA}
\author{V. Ho}
\affiliation{CLASSE, Cornell University, Ithaca, NY 14853, USA}
\author{G. Hoffstaetter}
\affiliation{CLASSE, Cornell University, Ithaca, NY 14853, USA}
\author{Matthias Liepe}
\email{mul2@cornell.edu}
\affiliation{CLASSE, Cornell University, Ithaca, NY 14853, USA}
\author{T. O'Connell}
\affiliation{CLASSE, Cornell University, Ithaca, NY 14853, USA}
\author{S. Posen}
\affiliation{CLASSE, Cornell University, Ithaca, NY 14853, USA}
\author{P. Quigley}
\affiliation{CLASSE, Cornell University, Ithaca, NY 14853, USA}
\author{J. Sears}
\affiliation{CLASSE, Cornell University, Ithaca, NY 14853, USA}
\author{V. Veshcherevich}
\affiliation{CLASSE, Cornell University, Ithaca, NY 14853, USA}

\author{A. Grassellino}
\affiliation{FNAL, Batavia, IL 60510, USA}
\author{A. Romanenko}
\affiliation{FNAL, Batavia, IL 60510, USA}
\author{D.A. Sergatskov}
\affiliation{FNAL, Batavia, IL 60510, USA}

\begin{abstract}
The superconducting RF linac for LCLS-II calls for 1.3 GHz 9-cell cavities with an average intrinsic quality factor $Q_0$ of 2.7\e{10} at 2.0~K and 16 MV/m accelerating gradient. Two niobium 9 cell cavities, prepared with nitrogen-doping at Fermilab, were assembled into the Cornell Horizontal Test Cryomodule (HTC) to test cavity performance in a cryomodule that is very similar to a full LCLS-II cryomodule. The cavities met LCLS-II specifications with an average quench field of 17 MV/m and an average $Q_0$ of 3\e{10}. The sensitivity of the cavities' residual resistance to ambient magnetic field was determined to be 0.5 n$\Omega$/mG during fast cool down. In two cool downs, a heater attached to one of the cavity beam tubes was used to induce large horizontal temperature gradients. Here we report on the results of these first tests of nitrogen-doped cavities in a cryomodule, which provide critical information for the LCLS-II project.
\end{abstract}

\maketitle

\section{Introduction}
The ``Linac Coherent Light Source-II'' Project will construct a 4 GeV CW superconducting linac in the first kilometer of the existing SLAC tunnel \cite{LCLS}. In order for economic feasibility in CW operation, superconducting RF cavities in the main linac must achieve an average intrinsic quality factor ($Q_0$) of 2.7\e{10} at 16 MV/m and 2.0 K. Nitrogen-doping of niobium cavities has been proposed to meet this high $Q_0$ specification \cite{Anna,HighQIPAC}. To test the feasibility of these goals, two 1.3 GHz International Linear Accelerator (ILC) shaped 9 cell cavities \cite{Haebel1992} were nitrogen-doped and welded into an ILC type liquid helium vessel at Fermilab (FNAL), placed in the Cornell Horizontal Test Cryomodule (HTC) and tested at Cornell. These tests will be referred to as HTC9-1 and HTC9-2 in the following. The HTC has previously been used to test a 7 cell Cornell ERL cavity \cite{NickHTC}. In this paper we discuss the results of these first nitrogen-doped 9 cell cavities tested in a cryomodule.

\section{Method and Cavity Preparation}
\begin{figure}
\includegraphics[scale=5.5]{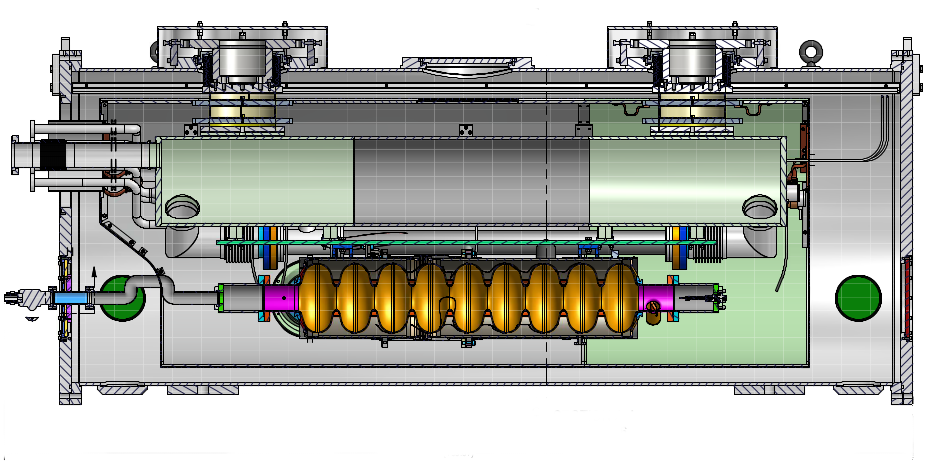}
\caption{A schematic of a 9-cell cavity in the Cornell HTC.}
\label{fig1}
\end{figure}

\begin{figure}
\centering
\includegraphics[scale=.05]{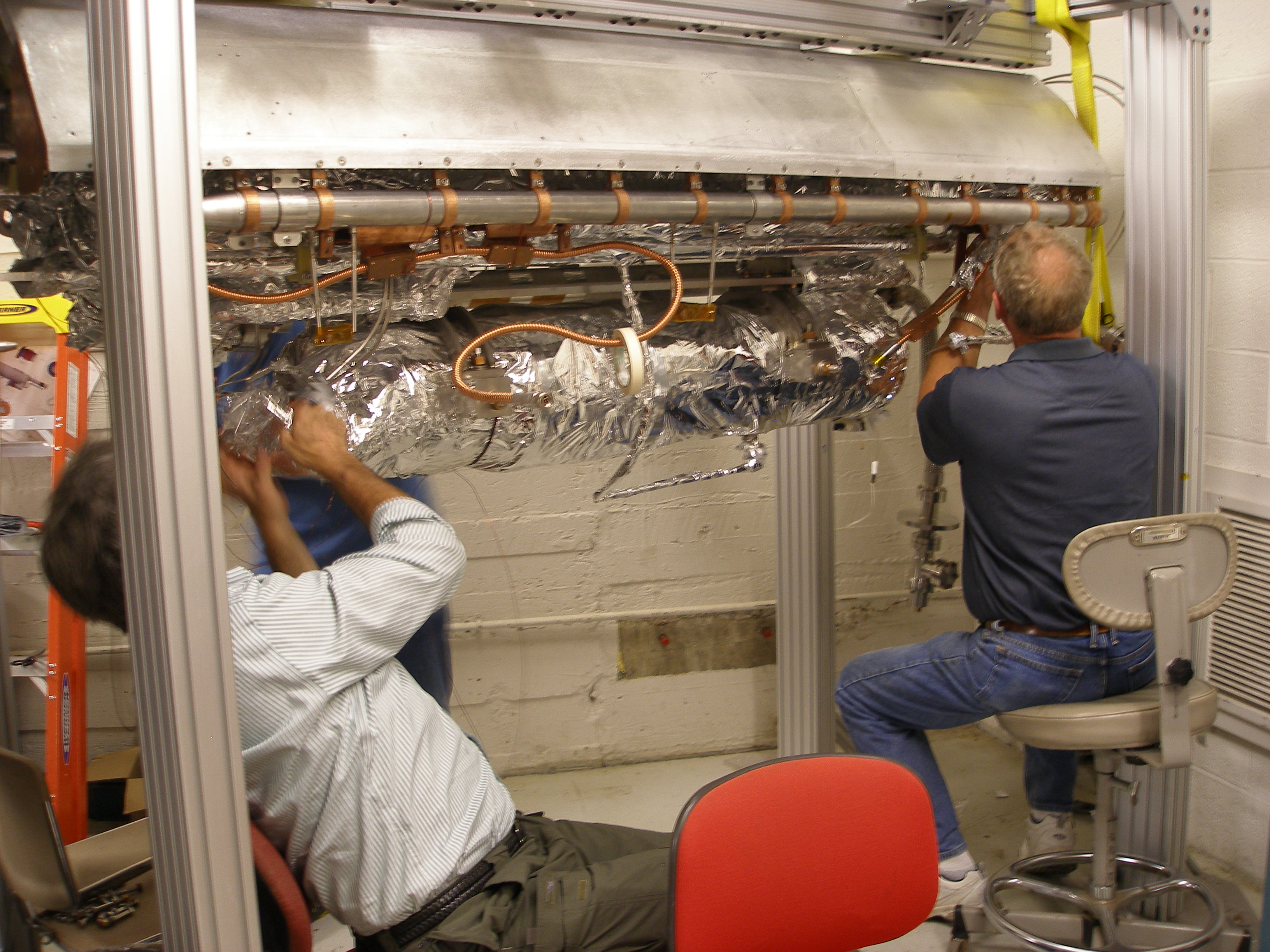}
\caption{The cavity being suspended from the HGRP.}
\label{fig2}
\end{figure}

\begin{figure}
\centering
\includegraphics[scale=.05]{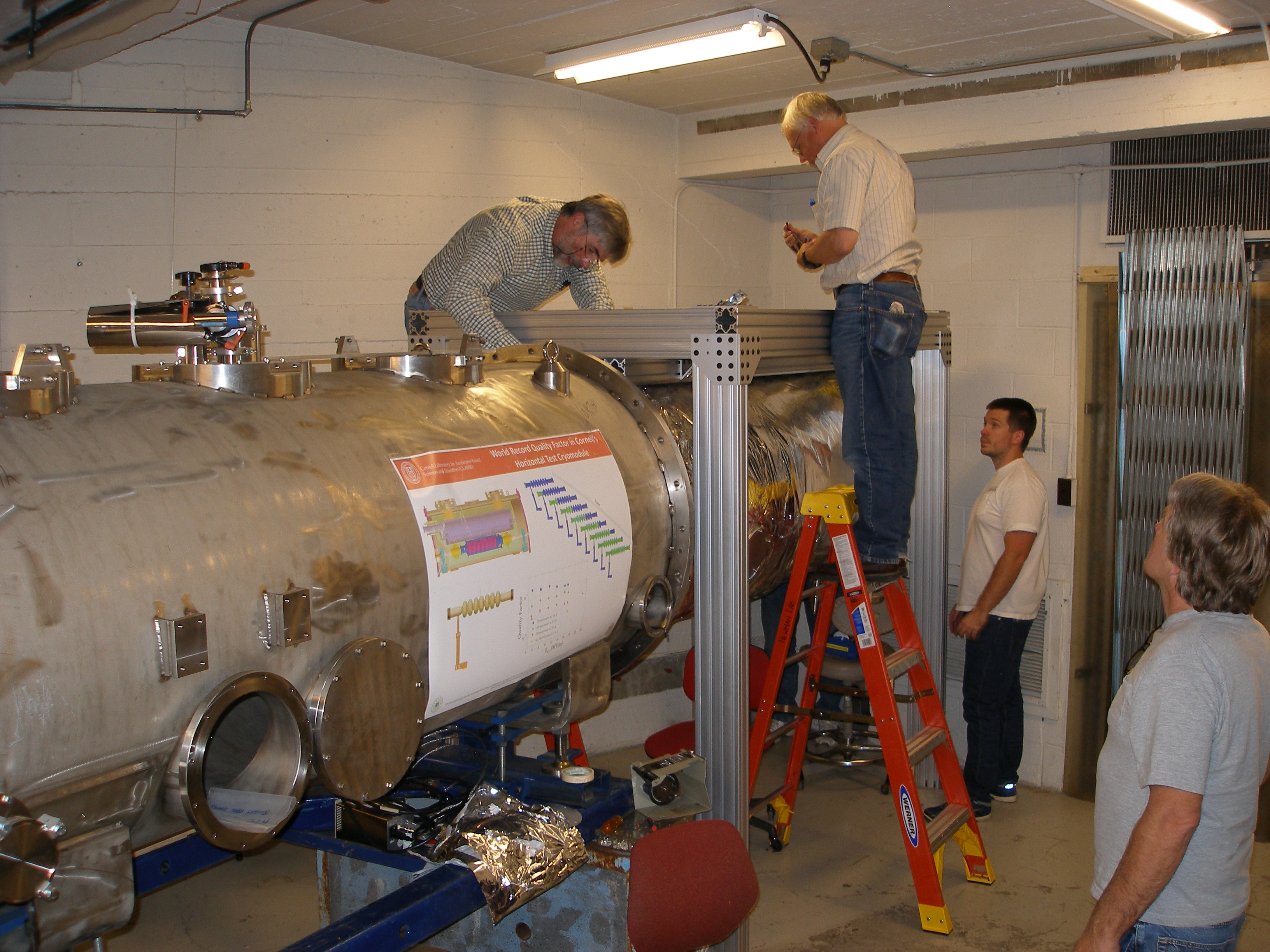}
\caption{The cryomodule cold-mass being slid into the vacuum vessel.}
\label{fig3}
\end{figure}

The Cornell HTC has been modified to hold a standard 9 cell ILC cavity. A schematic of the HTC with 9 cell cavity is shown in Fig.~\ref{fig1}. The HTC accepts a cavity that is dressed with helium tank and the cavity is suspended from the helium gas return pipe (HGRP) just as in a full cryomodule. For the two HTC tests discussed in this paper the cavities were welded into ILC style helium jackets with the helium input on one end (near the bottom of cell 2). The cross-section of the HTC is very similar to the proposed LCLS-II cryomodule and in practice represents a one-cavity section of that cryomodule. A high Q input coupler with $Q_{ext}\approx4\e{10}$ was used to excite fields in the cavities. The cavities (TB9AES012 and TB9AES011) were prepared with nitrogen-doping at FNAL. Typically, this consists of a bulk surface removal, heat treatment in vacuum, heat treatment in a nitrogen atmosphere, a heat treatment in vacuum again (called the soak stage), and a final electropolish (EP). Both cavities received a relatively high doping level. Because of the impact of doping level on cavity performance and sensitivity to magnetic fields, cavities with different levels of doping may have slightly different performance than those presented here. The specifics of the cavity preparations are shown in Table \ref{tab1}. Finally the cavity was dressed in a liquid helium tank at FNAL, and assembled in the cryomodule at Cornell. Figures \ref{fig2} and \ref{fig3} show the cavity being suspended from the HGRP and the cold-mass with the cavity being slid into the vacuum vessel, respectively.

\begin{table*}[tbh]
\centering
\begin{threeparttable}
\caption{HTC Cavity Preparation}
\begin{tabular}{l|x{5.5cm}x{5.5cm}}
\toprule
\textbf{Test} & HTC9-1 & HTC9-2 \\
\colrule
\textbf{Bulk Treatment} & CBP\tnote{1} and EP & EP \\
\textbf{Initial Heat Treatment} & 800$^\circ$C, 3 hours, vacuum & 800$^\circ$C, 90 minutes, vacuum \\
\textbf{N-Doping} & 25 mTorr, 20 minutes & 60 minutes\tnote{2} \\
\textbf{Soak Stage} & 30 minutes, vacuum & None \\
\textbf{Final EP} & 18 $\mu$m & 10 $\mu$m \\
\botrule
\end{tabular}
\label{tab1}
\begin{tablenotes}
\item[1] Centrifugal Barrel Polishing
\item[2] Mix of Argon and Nitrogen
\end{tablenotes}
\end{threeparttable}
\end{table*}

For both HTC tests, a variety of cool downs were completed. The purpose of these cool downs was to study the effect of cool down rate and spatial temperature gradients on the cavity performance. The exact details of the cool downs performed will be discussed in a later section. For each cool down, $Q_0$ vs $E_{acc}$ of the fundamental mode was measured at 1.6, 1.7, 1.8, 1.9, 2.0, and 2.1 K along with measurements of other TM010 modes at 2.0~K (and 1.6 K in the case of HTC9-2). Additionally, for one cool down in each test, $Q_0$ vs temperature from 4.2 to 1.6 K and resonance frequency vs temperature were measured. This was done in order to use BCS fitting with SRIMP \cite{DanHighQ, SRIMP} to extract BCS material properties.

\begin{figure}
\centering
\includegraphics[scale=.75]{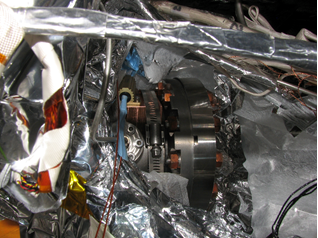}
\caption{A heater attached to the cell 9 beam tube to induce large longitudinal temperature gradients.}
\label{fig4}
\end{figure}

\begin{figure}
\centering
\includegraphics[scale=.75]{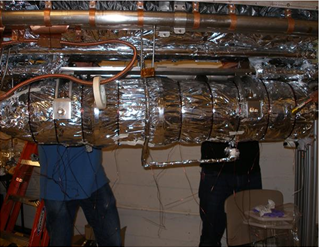}
\caption{The solenoid wrapped around the outside of the helium jacket to induce a uniform longitudinal external magnetic field.}
\label{fig5}
\end{figure}

\begin{figure*}
\centering
\includegraphics[scale=.75]{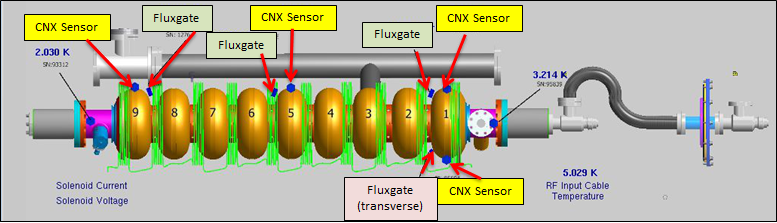}
\caption{The instrumentation for HTC9-2. Four Cernox sensors and four fluxgate magnetometers were used.}
\label{fig6}
\end{figure*}

In HTC9-2, a heater was placed on the cell 9 beam tube in order to induce a large longitudinal temperature gradient (Fig.~\ref{fig4}) and a solenoid was placed around the cavity's helium jacket to create a uniform external magnetic field parallel to the cavity axis (Fig.~\ref{fig5}). Figure \ref{fig6} shows a schematic of the cavity with instrumentation for HTC9-2. Four Cernox sensors were used  to measure temperature (one on the top of cell 1, 5, and 9, and one on the bottom of cell 1). Four fluxgate magnetometers were also placed in the same locations as the Cernox sensors, with the three on the top pointing along the cavity at a $45^\circ$ angle and the one on the bottom perpendicular to the cavity axis. In HTC9-1 there was an additional Cernox sensor placed on the bottom of cell 5 and there were only two fluxgates used, on either side of cell 5, one pointing parallel to the cavity axis and the other pointing perpendicular.

\section{Testing Overview}
For HTC9-1, four cool downs were completed: three fast and one slow. The first two fast cool downs were started at 80 K and the third at 150 K to induce large temperature gradients. Previous work on nitrogen-doped cavities has shown that fast cooling gives optimal performance and speculates that thermal gradients at the phase front may play a major role \cite{AnnaCoolDown}. Subsequent experimental work has verified this interpretation in vertical test \cite{DanLinacFlux,AlexMagField} and horizontal test \cite{DanLinacHTC}. For HTC9-2, seven cool downs were conducted: two fast from 150 K, one fast from 80 K, one slow from 20 K in an applied magnetic field, one fast from 150 K in an applied magnetic field, and two fast from 80 K using the heater to induce larger temperature gradients. Tables \ref{tab2} and \ref{tab3} show a summary of the cool downs and the various cooling parameter achieved in each.

\begin{table*}[tbh]
\centering
\begin{threeparttable}
\caption{HTC9-1 Cool Down Parameters}
\begin{tabular}{l|cccc}
\toprule
	\textbf{Parameter} & \textbf{Fast 1\tnote{1}} & \textbf{Fast 2} & \textbf{Slow 1} & \textbf{Fast 3}  \\
	\colrule
	Starting Temp [K] & 80 & 80 & 20 & 150  \\
	$|$dT/dt$|$ [K/min] & 1.4 & 6.8 & 0.7 & 2.4 \\
	Max $\Delta$T$_{\unit{horiz}}$ [K] &  0.5 & 0.2 & 0.8 & 2.0\\
	Max $\Delta$T$_{\unit{vert}}$ [K] & 7.9& 2.4 & 0.3 & 14.1 \\
	$|$B$_{\unit{long}}$(10 K)$|$ [mG] & 0.6 & 0.4 & 0.2 & 0.3  \\
	$|$B$_{\unit{perp}}$(10 K)$|$ [mG] & 0.3 & 0.1 & 0.1 & 1.5 \\
	$Q_0$(2.0~K, 14 MV/m) & 2.5\e{10} & 2.8\e{10} & 2.5\e{10} & 3.2\e{10} \\
	R$_{\unit{res}}$ (14 MV/m) [n$\Omega$] &  $5\pm1$ & $4.0\pm0.8$ & $5\pm1$ & $2.7\pm0.5$ \\
	\botrule
\end{tabular}
\begin{tablenotes}
\item[1] After field emission conditioning
\end{tablenotes}
\label{tab2}
\end{threeparttable}
\end{table*}

\begin{table*}[tbh]
\centering
\caption{HTC9-2 Cool Down Parameters}
\begin{tabular}{l|ccccccc}
\toprule
	\textbf{Parameter} & \textbf{Fast 1} & \textbf{Fast 2} & \textbf{Fast 3} & \textbf{Fast 4} & \textbf{Slow 1} & \textbf{Fast 5} & \textbf{Fast 6} \\
	\colrule
	Solenoid On & No & No & No & Yes & Yes & No & No \\
	Heater Power [W] & 0 & 0 & 0 & 0 & 0 & 50 & 100 \\
	Starting Temp [K] & 150 & 150 & 80 & 150 & 20 & 80 & 80 \\
	$|$dT/dt$|$ [K/min] & 1.2 & 1.3 & 2.2 & 1.3 & 0.4 & 4.7 & 4.9 \\
	Max $\Delta$T$_{\unit{horiz}}$ [K] & 5.2 & 6.8 & 9.4 & 5.9 & 0.2 & 21.8 & 29.4 \\
	Max $\Delta$T$_{\unit{vert}}$ [K] & 6.2 & 8.9 & 16.1 & 7.0 & 0.5 & 32.5 & 39.4 \\
	$|$B$_{\unit{long}}$(10 K)$|$ [mG] & 1.2 & 1.2 & 4.4 & 10.1 & 10.6 & 2.5 & 3.1  \\
	$|$B$_{\unit{perp}}$(10 K)$|$ [mG] & 6.3 & 0.3 & 3.1 & 7.2 &1.0 & 50.3 & 62.8 \\
	$Q_0$(2.0~K, 16 MV/m) & 2.6\e{10} & 2.7\e{10} & 2.4\e{10} &1.9\e{10} & 6.2\e{9} & 2.1\e{10} & 1.5\e{10} \\
	R$_{\unit{res}}$ (16 MV/m) [n$\Omega$] & $5.2\pm1$ & $4.5\pm0.9$ & $4.6\pm0.9$ & $9\pm1.4$ & $40\pm8$ & $7.3\pm0.9$ & $12\pm2$ \\
	\botrule
\end{tabular}
\label{tab3}
\end{table*}

\section{Cool Down Dynamics}
\begin{figure}
\centering
\includegraphics[scale=.27]{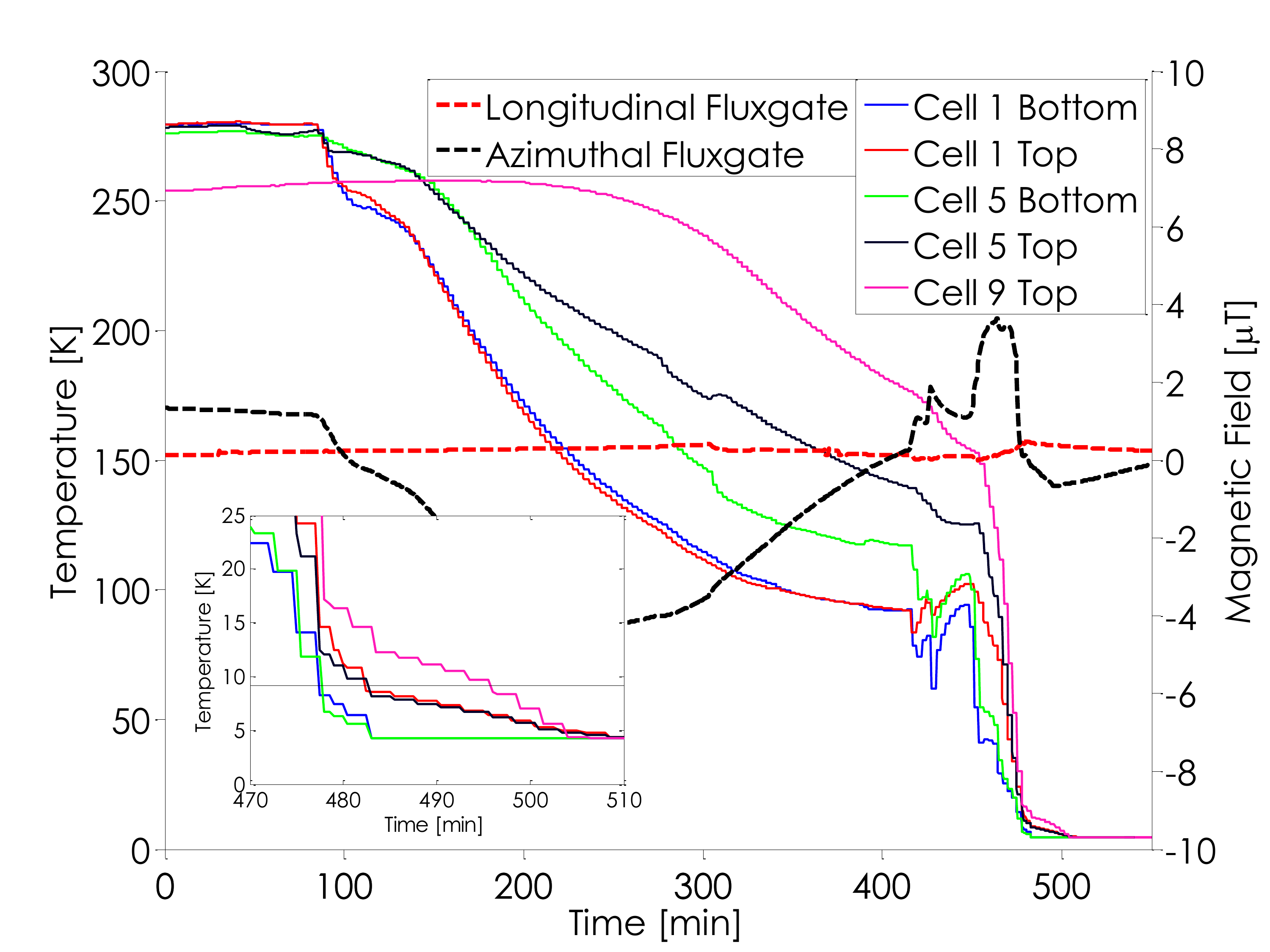}
\caption{Temperature and magnetic field profile during cooling for the third fast cool down of HTC9-1.}
\label{fig7}
\end{figure}

\begin{figure}
\centering
\includegraphics[scale=.33]{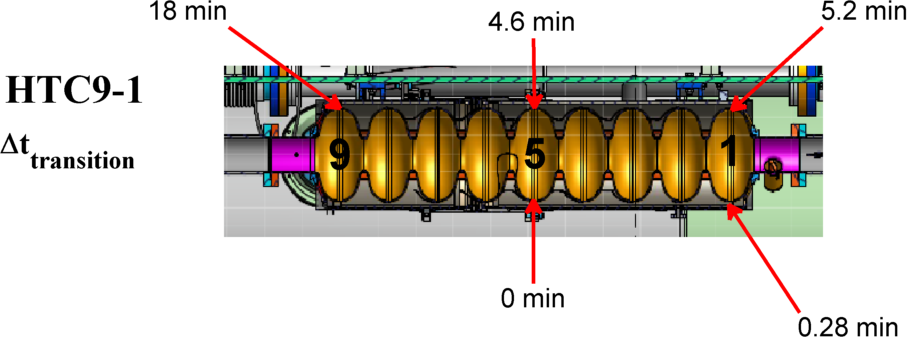}
\caption{A schematic of the cavity showing how the cavity transitions from the normal-conducting to the superconducting state (shown for the third fast cool down of HTC9-1).}
\label{fig8}
\end{figure}

In all cool downs, cooling was done by filling with liquid helium from the bottom of the liquid helium tanks. The cryogenic input port is located near the bottom of cell 2 on the helium jacket. As an example, the temperatures and magnetic field values of the third fast cool down from HTC9-1 are shown in Fig.~\ref{fig7} as a function of time. Figure~\ref{fig8} shows the time evolution of the cavity's transitioning to the superconducting state  during this cool down. Each temperature sensor location is marked by the time interval between that sensor going through T$_c$ and the first sensor going through T$_c$. The cool down data from the fast cool downs consistently  shows  that the bottom of the cavity cools first. The top of cell 5 follows shortly after and then the top of cell 1. The top of cell 9 cools last, significantly later. Liquid begins accumulation very quickly on the bottom of the tank, resulting in the very fast cooling of the bottom. Due to the location of the chimney (the top of cell 3), gas preferentially flows towards that end, resulting in the tops of cells 1-5 cooling faster than the tops of cells 6-9. 

During fast cool down in the HTC, we were able to achieve rates on the order of 3 K/min, similar to what is typically done in fast vertical cavity performance test. In the fast HTC cool downs, vertical spatial temperature gradients achieved are 10 to 15 K (corresponding to 50 to 75 K/m), smaller than what can be achieved in vertical cavity performance test where spatial gradients more than 100 K/m are typically reached in fast cool down. These large gradients are important for flux expulsion which will be discussed later. Horizontal temperature gradients are less than 10 K, small enough to minimize magnetic fields from thermoelectric currents, also to be discussed later.

\section{$Q_0$ vs $E_{acc}$ Performance}
\begin{figure}
\centering
\includegraphics[scale=.27]{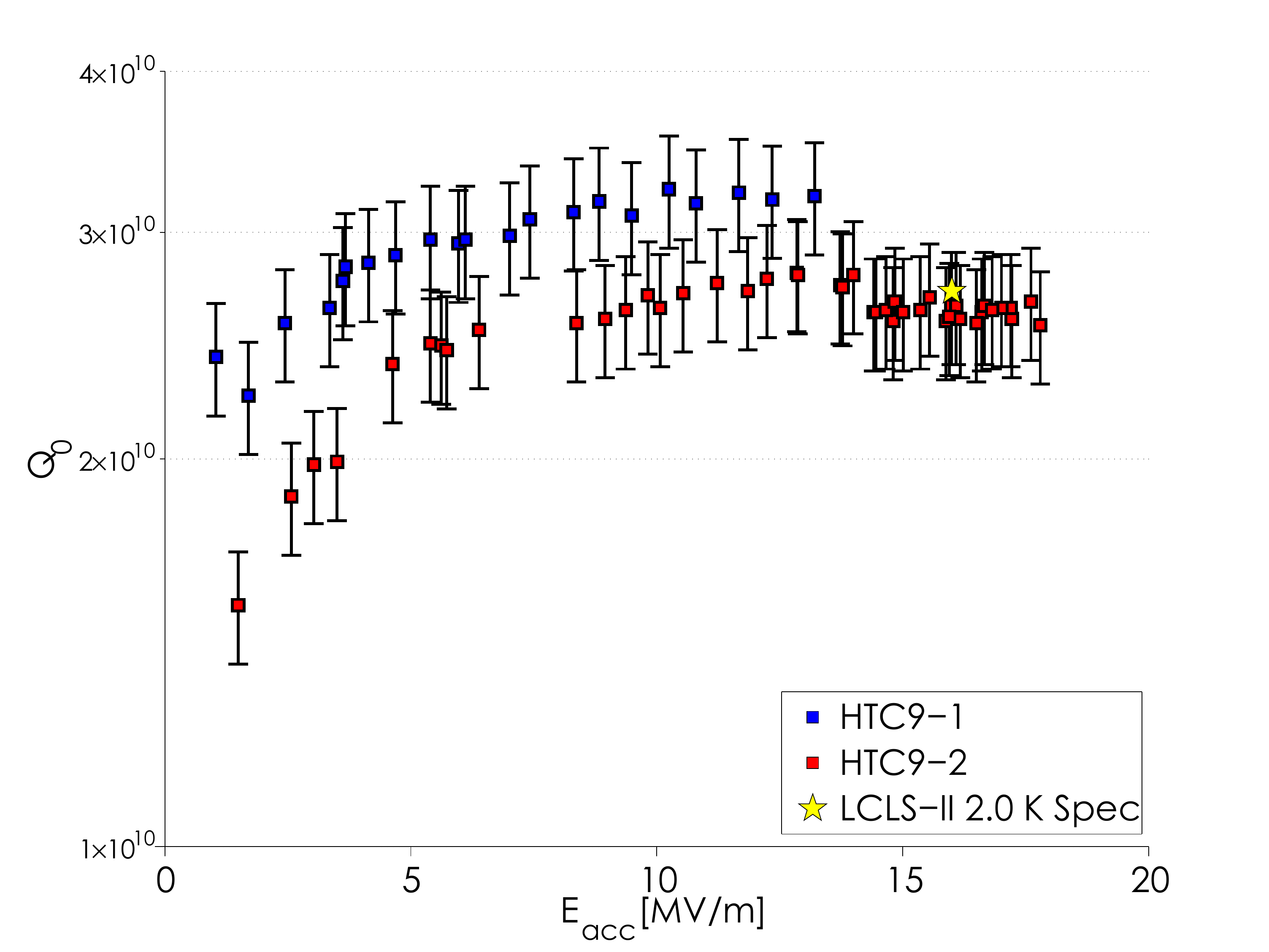}
\caption{The best $Q_0$ vs $E_{acc}$ performance for the two HTC tests. Uncertainty on $E_{acc}$ is  10\%.}
\label{fig9}
\end{figure}

The dynamics of a given cool down had a strong impact on the cavity performance in both HTC9-1 and HTC9-2. Figure \ref{fig9} shows the best performance for the two tests at 2.0 K. HTC9-2 meets LCLS-II specifications for both $Q_0$ and $E_{acc}$ while HTC9-1 meets and significantly exceeds the LCLS-II $Q_0$ specification but has a quench field slightly lower than the $E_{acc}$ specification. On average the two cavities reached a $Q_0$ of 3\e{10} and a maximum $E_{acc}$ of 17 MV/m. These results are very promising for the LCLS-II project.

\subsection{HTC9-1}
\begin{figure}
\centering
\includegraphics[scale=.27]{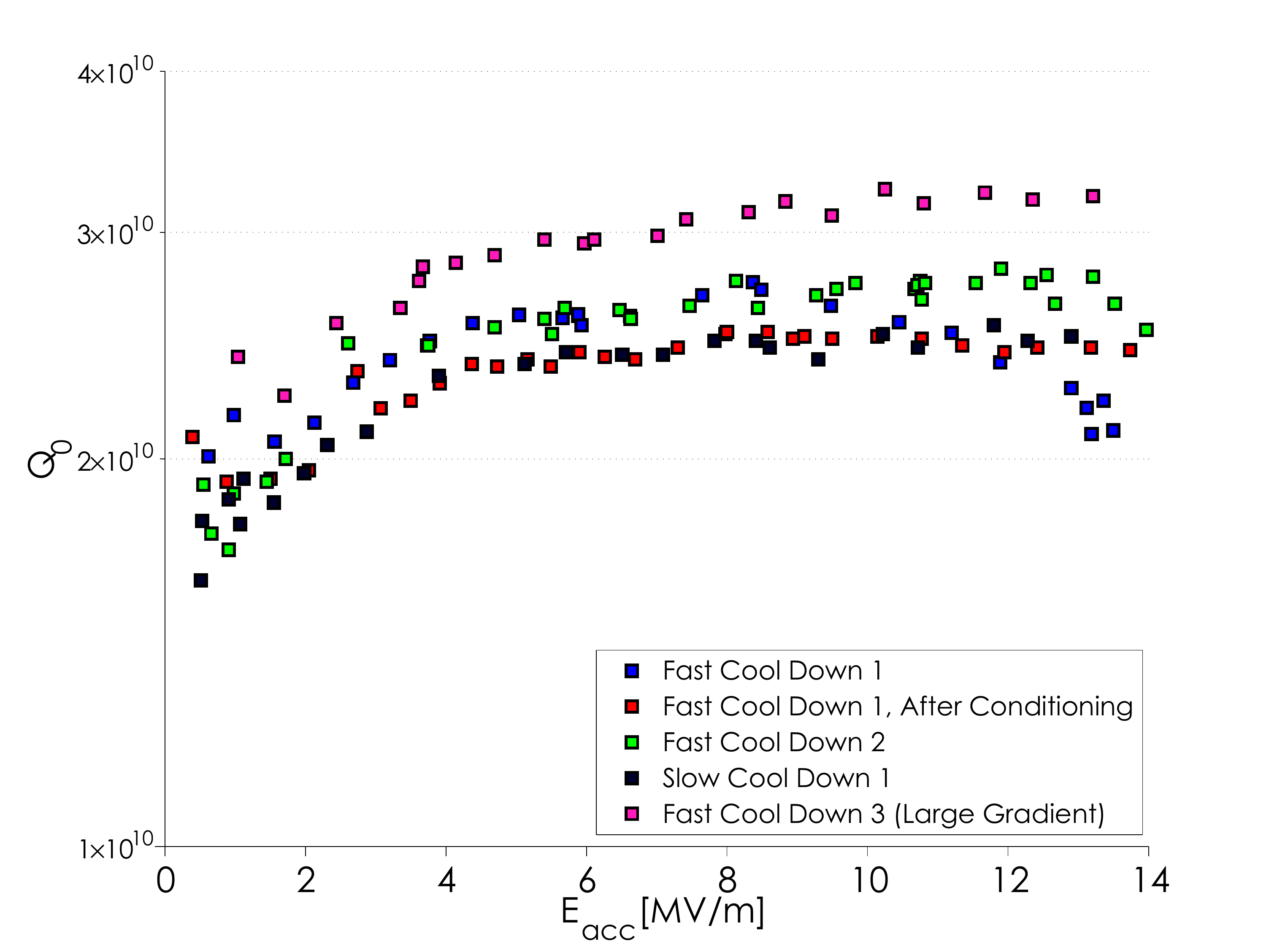}
\caption{The 2.0 K $Q_0$ vs $E_{acc}$ performance for all cool downs for HTC9-1. Uncertainty on $E_{acc}$ and $Q_0$ are  10\%.}
\label{fig10}
\end{figure}

\begin{figure}
\centering
\includegraphics[scale=.27]{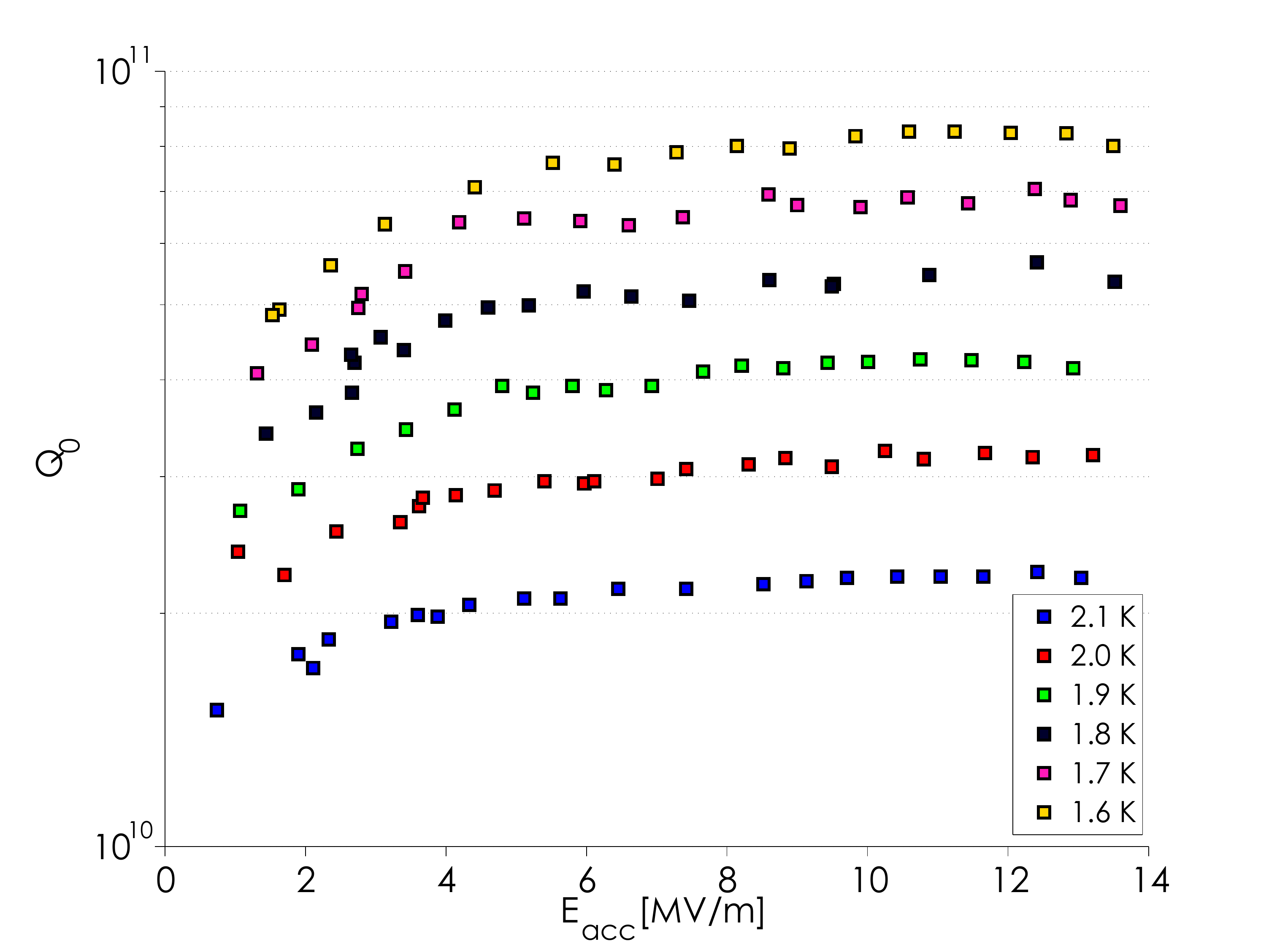}
\caption{The $Q_0$ vs $E_{acc}$ performance at all temperatures for the third fast cool down of HTC9-1. Uncertainty on $E_{acc}$ and $Q_0$ are  10\%.}
\label{fig11}
\end{figure}

In HTC9-1, the cavity quenched at $14\pm1$ MV/m. Comparatively, in vertical test the cavity quenched at $15\pm1$ MV/m. Between vertical and horizontal test, the cavity was re-tuned for field flatness which can have an impact on the quench field on the $\pm$5\% level. It is therefore clear that moving to horizontal test did not significantly alter the quench field. The $Q_0$ was strongly dependent on the cool down, as can be seen in Fig.~\ref{fig10}, which shows the 2.0~K $Q_0$  for all four cool downs. The third fast cool down, which had the largest vertical temperature gradients, produced the best results with a $Q_0$ of 3.2\e{10} at 14 MV/m and 2.0~K. The $Q_0$ vs $E_{acc}$ performance at all temperatures measured for that cool down are shown in Fig.~\ref{fig11}. A more detailed analysis of the performance of HTC9-1 is presented in \cite{DanLinacHTC}. Initially the 2.0~K $Q_0$  was 2.5\e{10} at maximum fields. This lower $Q_0$  was a result of conditioning field emission of the cavity in which the cavity was quenched many times. After a thermal cycle, and similar fast cool down, the 2.0~K $Q_0$  improved to 2.8\e{10} at 14 MV/m. A slow cool down resulted in a lower $Q_0$  of 2.5\e{10}, consistent with previous observation during vertical performance tests of nitrogen-doped cavities. The final fast cool down, in which larger vertical gradients were achieved than in the first two fast cool downs resulted in a 2.0~K $Q_0$  of 3.2\e{10} at 14 MV/m, significantly exceeded LCLS-II $Q_0$  specification. Additionally, the $Q_0$ at 1.6 K was 8\e{10} for the final fast cool down (Fig.~\ref{fig11}).

\subsection{HTC9-2}
\begin{figure}
\centering
\includegraphics[scale=.27]{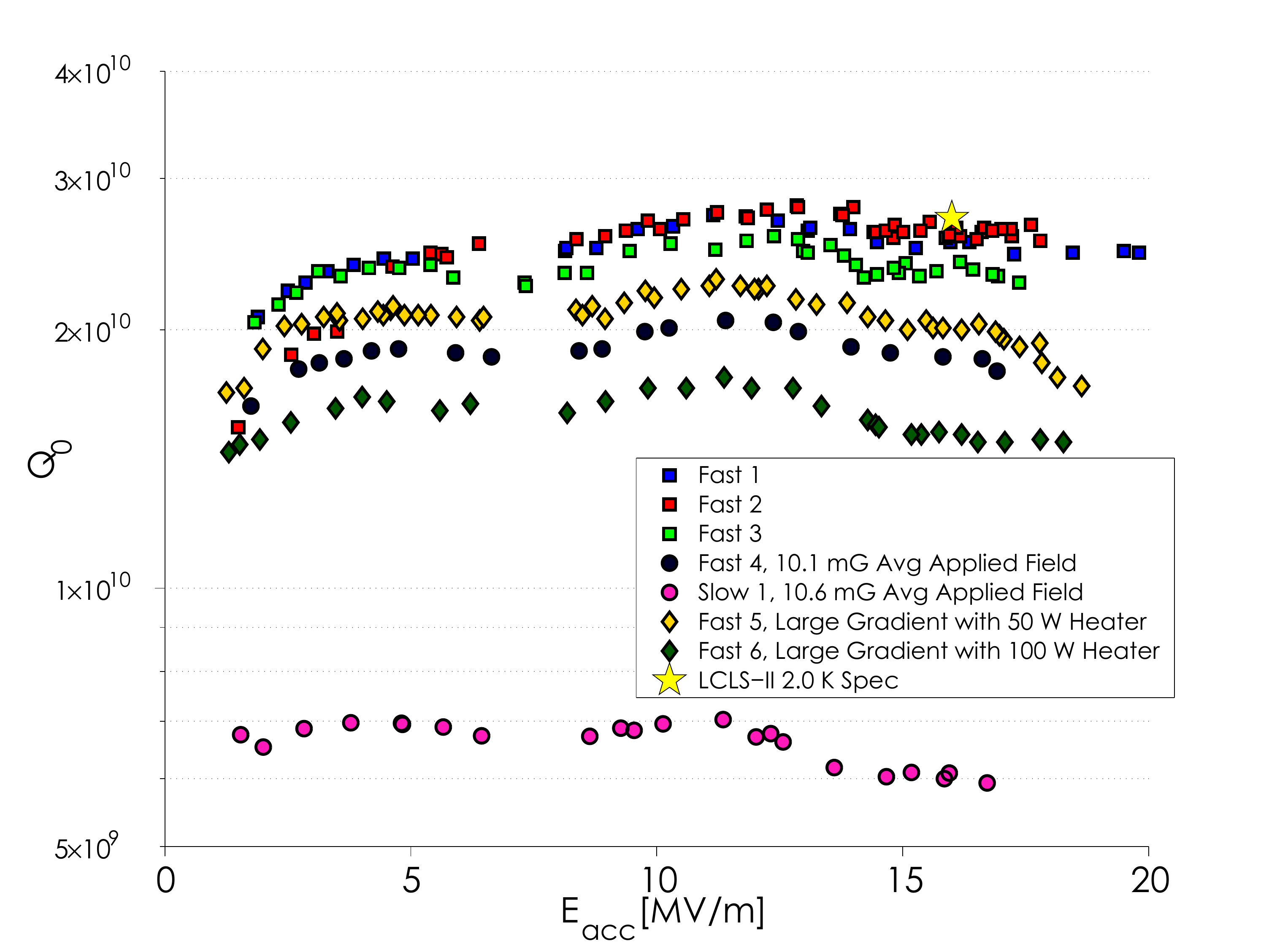}
\caption{The 2.0 K $Q_0$ vs $E_{acc}$ for HTC9-2 for all seven cool downs. Best performance was achieved in the second fast cool down. Uncertainty on $E_{acc}$ and $Q_0$ are  10\%.}
\label{fig12}
\end{figure}

\begin{figure}
\centering
\includegraphics[scale=.27]{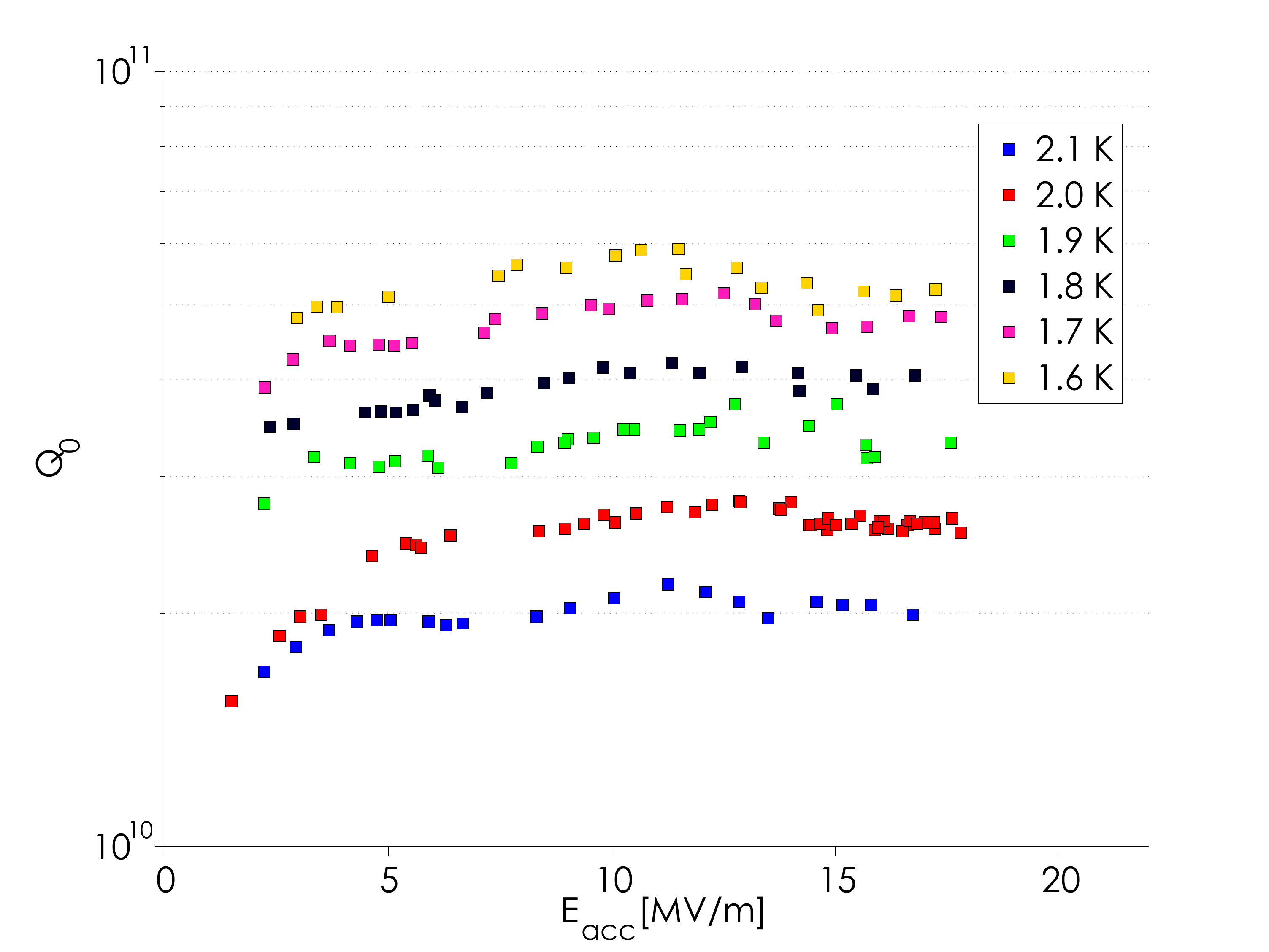}
\caption{$Q_0$ vs $E_{acc}$ for all measured temperatures for the second fast cool down of HTC9-2. Uncertainty on $E_{acc}$ and $Q_0$ are  10\%.}
\label{fig13}
\end{figure}

In HTC9-2, the cavity quenched at $20\pm2$ MV/m, consistent with maximum fields during vertical cavity performance test. $Q_0$ was again highly dependent on the cool down dynamics. Figure~\ref{fig12} shows the 2.0~K performance for the cavity in all seven cool downs. The best results came from the second fast cool down. In this cool down the $Q_0$ was 2.7\e{10} at 2.0~K and 16~MV/m, meeting the LCLS-II specification. The cool downs with applied ambient magnetic field and with the beam tube heater, and how the $Q_0$ was impacted, will be discussed in later sections. Figure \ref{fig13} shows the $Q_0$ vs $E_{acc}$ performance of the cavity at all temperatures for the second fast cool down which resulted in the best performance. 

\begin{figure}
\centering
\includegraphics[scale=.27]{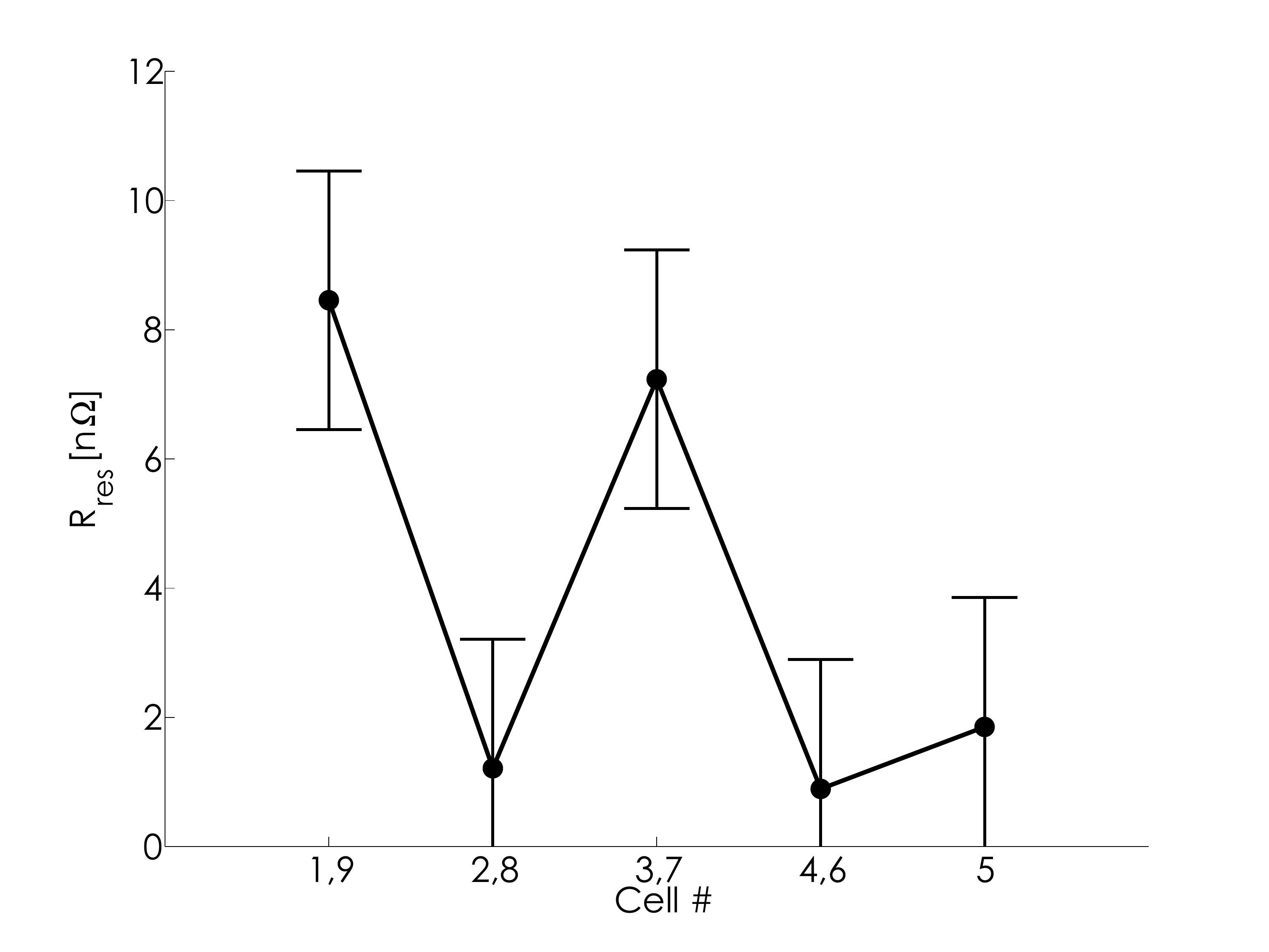}
\caption{The residual resistance distribution by cell for the second fast cool down of HTC9-2.}
\label{fig20}
\end{figure}

From measurements of other TM010 modes at 1.6 and 2.0 K, surface resistance and thus residual resistance can be extracted for each cell pair (cells 1 and 9, 2 and 8, 3 and 7, 4 and 6, and 5). Figure \ref{fig20} shows the residual resistance distribution for the second fast cool down, demonstrating that residual resistance is not uniformly distributed among the cells. 

\subsection{Changes from Vertical to Horizontal Test}
It is interesting to note the changes in $Q_0$ from vertical test at FNAL to horizontal test at Cornell. Table~\ref{tab4} shows a summary of the changes that took place. These changes came from a difference in residual resistance, most likely caused by the higher ambient magnetic field in the HTC and differences in the cool down dynamics. In vertical test, cavities can be cooled with such large gradients and ambient magnetic fields are so small that residual resistance from the magnetic field is minimal. The dependence of residual resistance on cool down dynamics is presented thoroughly in \cite{AnnaCoolDown}, while the dependence on magnetic field trapped is discussed in \cite{DanLinacFlux}. This data represents the first measurements of performance degradation for nitrogen doped cavities from vertical to horizontal test. These first two tests suggest an increase of $\sim$1-3 n$\Omega$ of residual resistance between vertical and horizontal test in a 2 mG ambient magnetic field. Since the magnetic field in the HTC is $\sim$2 mG,  this increase is consistent with $\sim$1 n$\Omega$/mG of additional residual resistance for a given magnetic field using  the Cornell Fast Cool Down Procedure.

\begin{table}[tbh]
\centering
\caption{A summary of the change in $Q_0$ from vertical to horizontal test for the two cavities in their best cool down}
\begin{tabular}{l|ccc}
\toprule
\textbf{Cavity} & \textbf{Vertical $Q_0$} & \textbf{Horizontal $Q_0$} & \textbf{$\Delta$R$_{\unit{res}}$ [n$\Omega$]} \\
\colrule
HTC9-1 & 3.5\e{10} & 3.2\e{10} & $\sim1$ \\
HTC9-2 & 3.4\e{10} & 2.7\e{10} & $\sim3$ \\
\botrule
\end{tabular} 
\label{tab4}
\end{table}

\section{Material Properties}
By measuring resonance frequency as a function of temperature during warm-up and $Q_0$ vs temperature between 4.2 and 1.6 K, material properties such as $T_c$, energy gap ($\Delta/k_BT_c$), and mean free path were determined for the two cavities. Change in resonance frequency is converted to change in penetration depth with the method discussed in \cite{DanIPACLCLS,Kneisel,Visentin,GiGiBaking} and is used to fit for $T_c$ and mean free path. The $Q_0$ vs T data is fit for energy gap and residual resistance. A summary of the material properties obtained  is shown in Table \ref{tab5} for the two cavities. Notice the small values for the mean free path, which are the result of the strong nitrogen doping of the RF surface layer during the heat treatment in nitrogen atmosphere.

\begin{table}
\centering
\caption{Material properties extracted at 5 MV/m for HTC9-1 and HTC9-2}
\begin{tabular}{l|cc}
\toprule
\textbf{Material Property} & \textbf{HTC9-1} & \textbf{HTC9-2} \\
\colrule
T$_c$ [K] & $9.2\pm0.2$ & $9.2\pm0.2$ \\
$\Delta/k_BT_c$ & $1.94\pm0.02$ & $1.94\pm0.02$ \\
Mean Free Path [nm] & $58\pm17$ & $29\pm9$ \\
\botrule
\end{tabular}
\label{tab5}
\end{table}

\section{Field Dependence of the Surface Resistance}
\begin{figure}
\centering
\includegraphics[scale=.27]{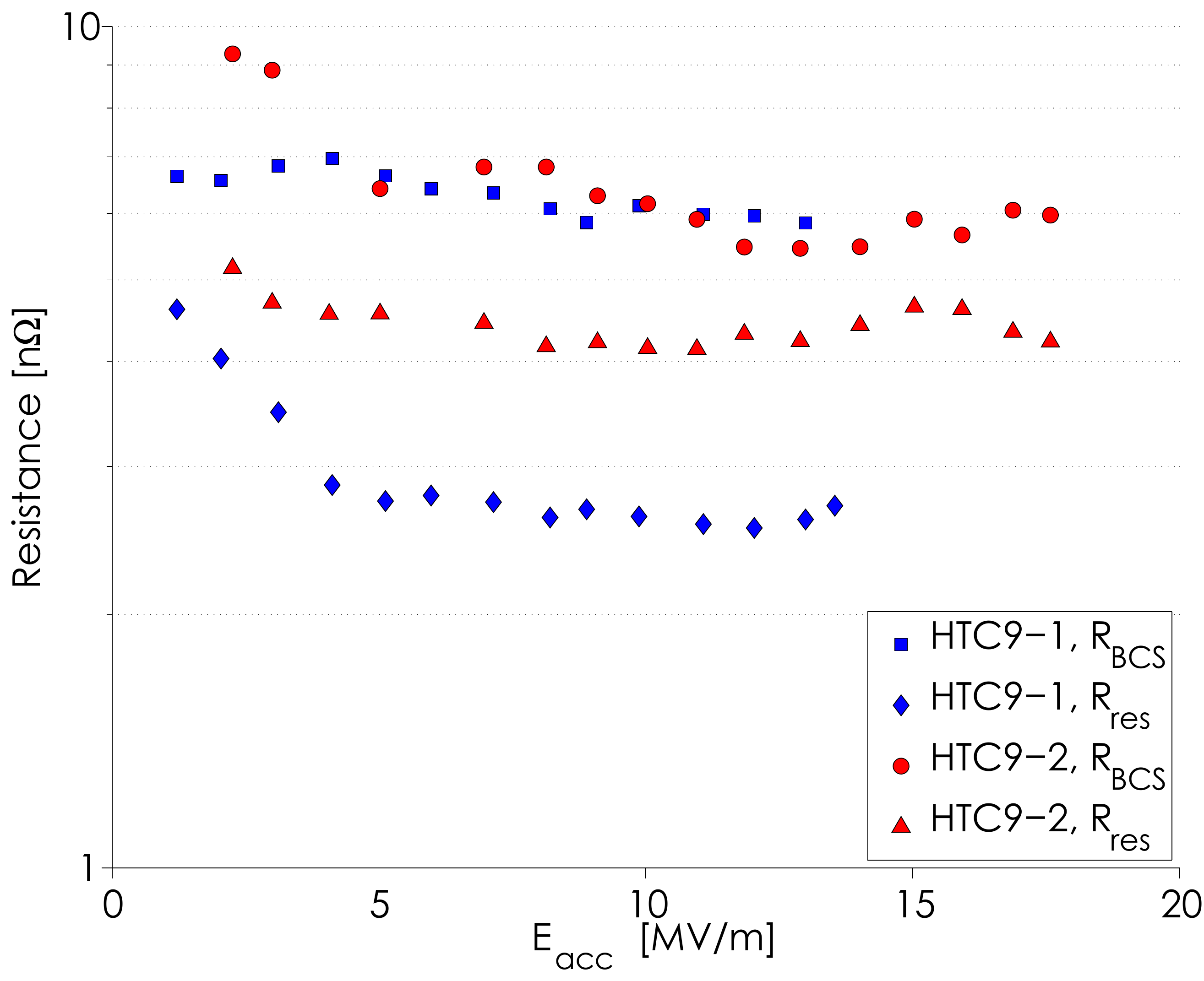}
\caption{The 2.0 K BCS resistance and residual resistance for the best cool downs of HTC9-1 and HTC9-2.}
\label{fig14}
\end{figure}

BCS fitting can also be applied to field dependent data to find the field dependence of the residual and BCS resistances \cite{DanN2,Alex}.  Examining this field dependence reveals  that the anti-Q slope observed is caused by a decreasing BCS resistance, consistent with previous measurements \cite{Anna,DanN2}. This can be seen clearly in Fig.~\ref{fig14}, which shows the 2.0~K BCS resistance and residual resistance for the best cool downs of HTC9-1 and HTC9-2 as a function of accelerating field. For both cavities, the BCS resistance decreases from about 8 to 6 n$\Omega$ in the medium field region. Meanwhile, the residual resistance is constant above the low field Q slope region. It has been shown in previous work that the BCS resistance approximately decreases with the logarithm of the field for nitrogen-doped cavities \cite{GiGiLog,DanLinacLCLS}.

\section{Conditioning of Field Emission in the 8$\pi$/9 Mode}
\begin{figure}
\centering
\includegraphics[scale=.27]{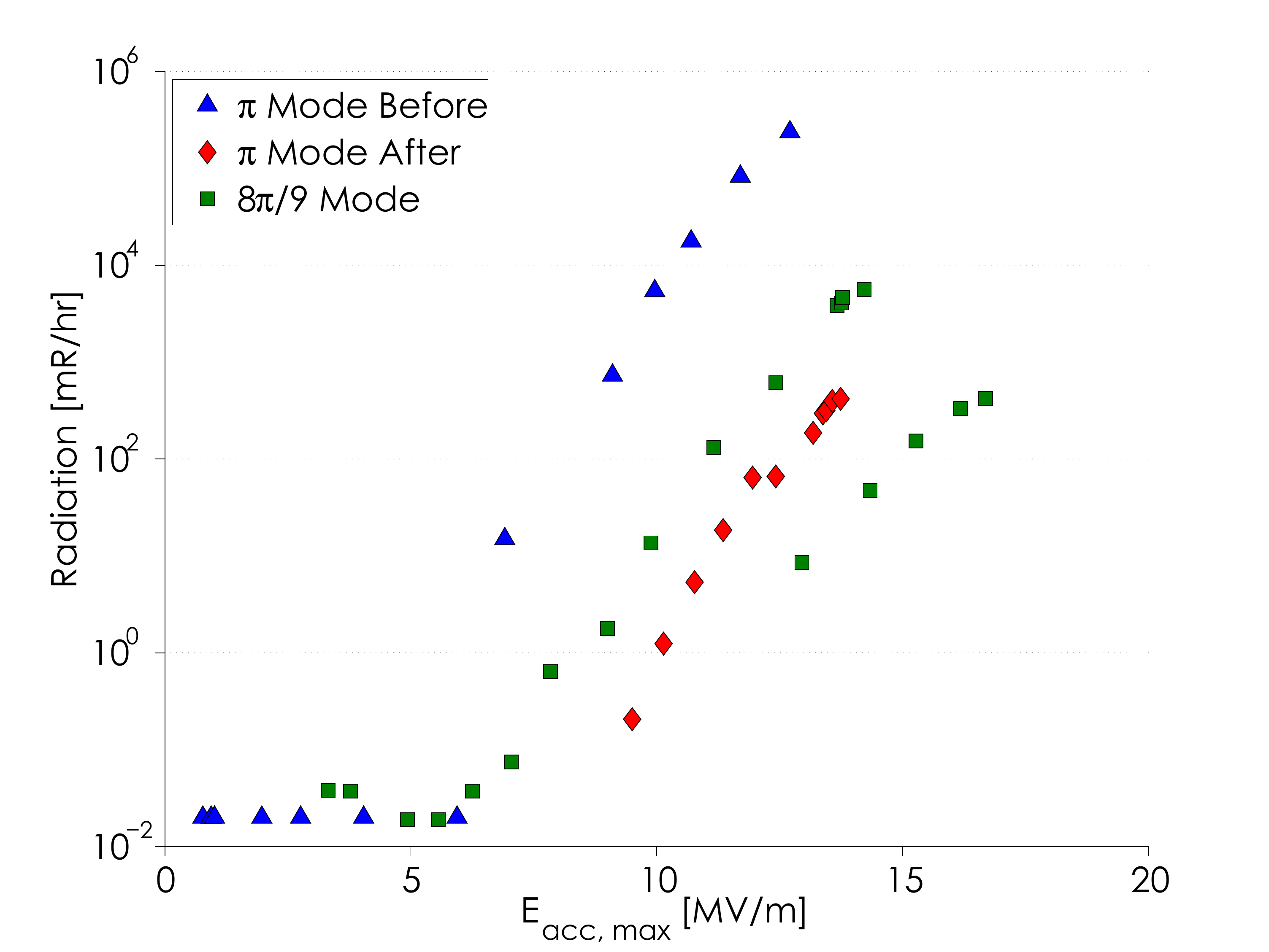}
\caption{Field emission conditioning in the 8$\pi$/9 mode for HTC9-1. Radiation in the $\pi$ mode is also shown before and after conditioning.}
\label{fig15}
\end{figure}

During the first test of HTC9-1, radiation was very high, $\sim$100 R/hr at 13.5 MV/m. The field emitter was suspected to be in the end cells and by driving the cavity in the 8$\pi$/9 mode, higher fields in the end cells could be reached than in the fundamental mode,  in which fields were limited to $\sim$14 MV/m in all cells by quench in one of the cells. The higher end-cell fields in the 8$\pi$/9 mode were able to condition the field emitter. This is shown in Fig.~\ref{fig15}, which shows the radiation as a function of maximum accelerating field for the 8$\pi$/9 mode and the $\pi$ mode before and after conditioning. The radiation in the $\pi$ mode is reduced by a factor of 1000 after conditioning. This demonstrates a powerful method for conditioning field emission in multicell cavities by driving the cavity in a different TM010 mode. Additionally, because the quench field of the fundamental mode was unchanged after conditioning, this confirms that the quench was not caused by field emission, but by some unrelated other type of defect.

\section{$Q_0$ Dependence on Beam Tube Temperature}
\begin{figure}
\centering
\includegraphics[scale=.27]{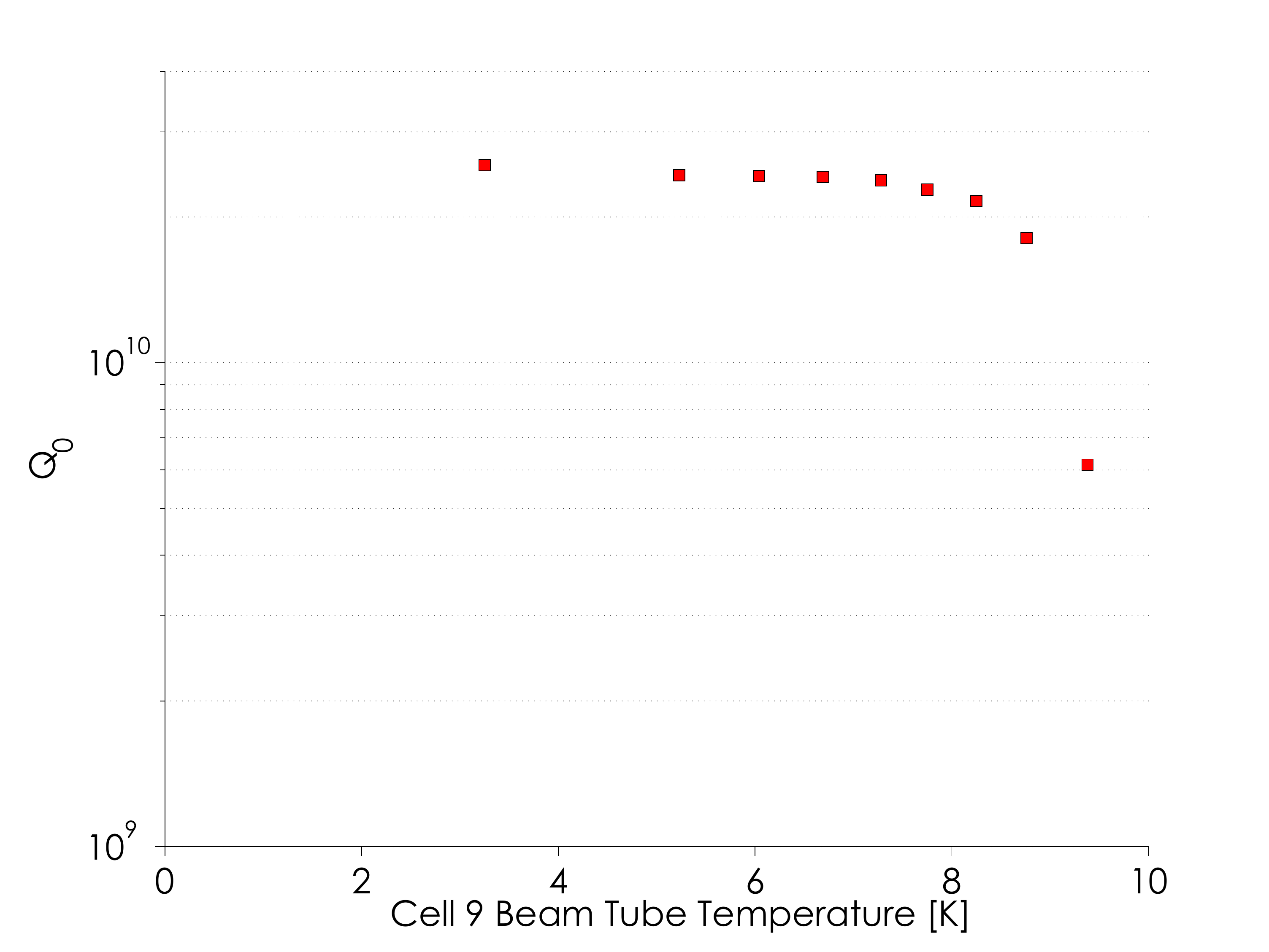}
\caption{The $Q_0$ at 2.0 K and 10 MV/m as a function of beam tube temperature for HTC9-2. Uncertainty on  $Q_0$ is  10\%.}
\label{fig16}
\end{figure}

\begin{figure}
\centering
\includegraphics[scale=.27]{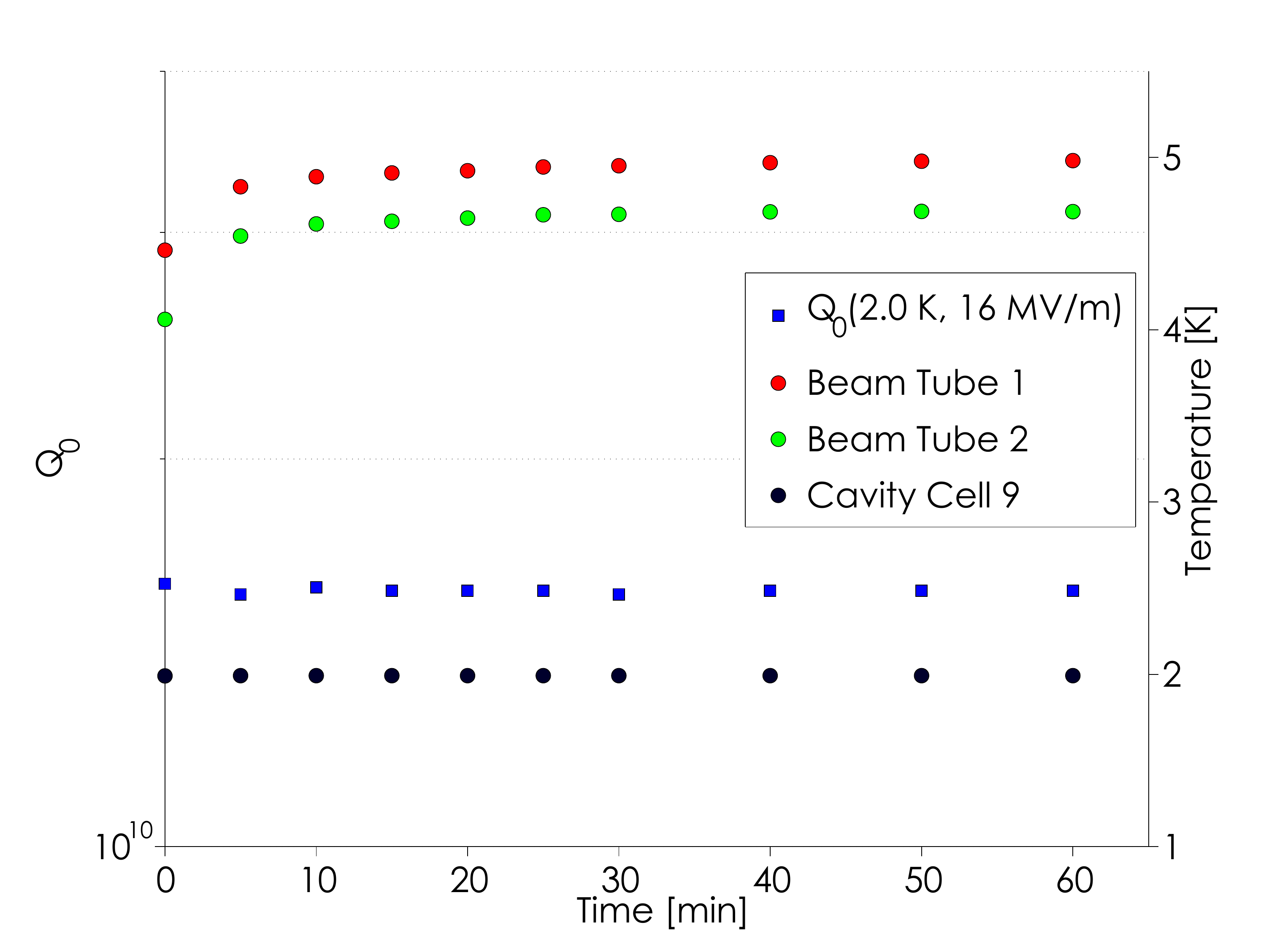}
\caption{The $Q_0$ at 2.0 K and 16 MV/m as a function of time with the cavity at field in HTC9-2. Also shown are the two beam tube temperatures and cavity temperature as a function of time. Uncertainty on  $Q_0$ is  10\%.}
\label{fig17}
\end{figure}

During the fifth fast cool down of HTC9-2, $Q_0$ was measured as a function of beam tube temperature. Using the heater attached to the beam tube on cell \#9, the beam tube's temperature was increased. $Q_0$ at 2.0 K and 10 MV/m was measured to check the dependence of $Q_0$ on beam tube temperature. The results are shown in Fig.~\ref{fig16}. The results show that $Q_0$ is unaffected by the beam tube temperature until it reaches 7 K where $Q_0$ begins to drop substantially. In order to explore if this effect would have a strong impact on $Q_0$ during CW cavity operation in LCLS-II the $Q_0$ at 16 MV/m and 2.0 K (operating temperature and gradient) was measured as a function of time that the cavity was at field. These results are shown in Fig.~\ref{fig17}. Also shown are the two beam tube temperatures and the cavity temperature. The cavity temperature was unaffected. Initially the beam tubes increased in temperature by $\sim$0.5 K and then stabilized around 5 K.  The data shows that  $Q_0$ of the cavity was not significantly impacted by this temperature increase, in agreement with the results of the measurements of $Q_0$ as a function of beam tube temperature. These results demonstrate that  end-group heating in continuous LCLS-II cavity operation can be kept low enough so as to not significantly reduce  the high  $Q_0$  of nitrogen-doped cavities. These measurements should be repeated with the high power RF input coupler and HOM pickup antennas in place.

\section{Solenoid Study on HTC9-2}
\begin{figure}
\centering
\includegraphics[scale=.27]{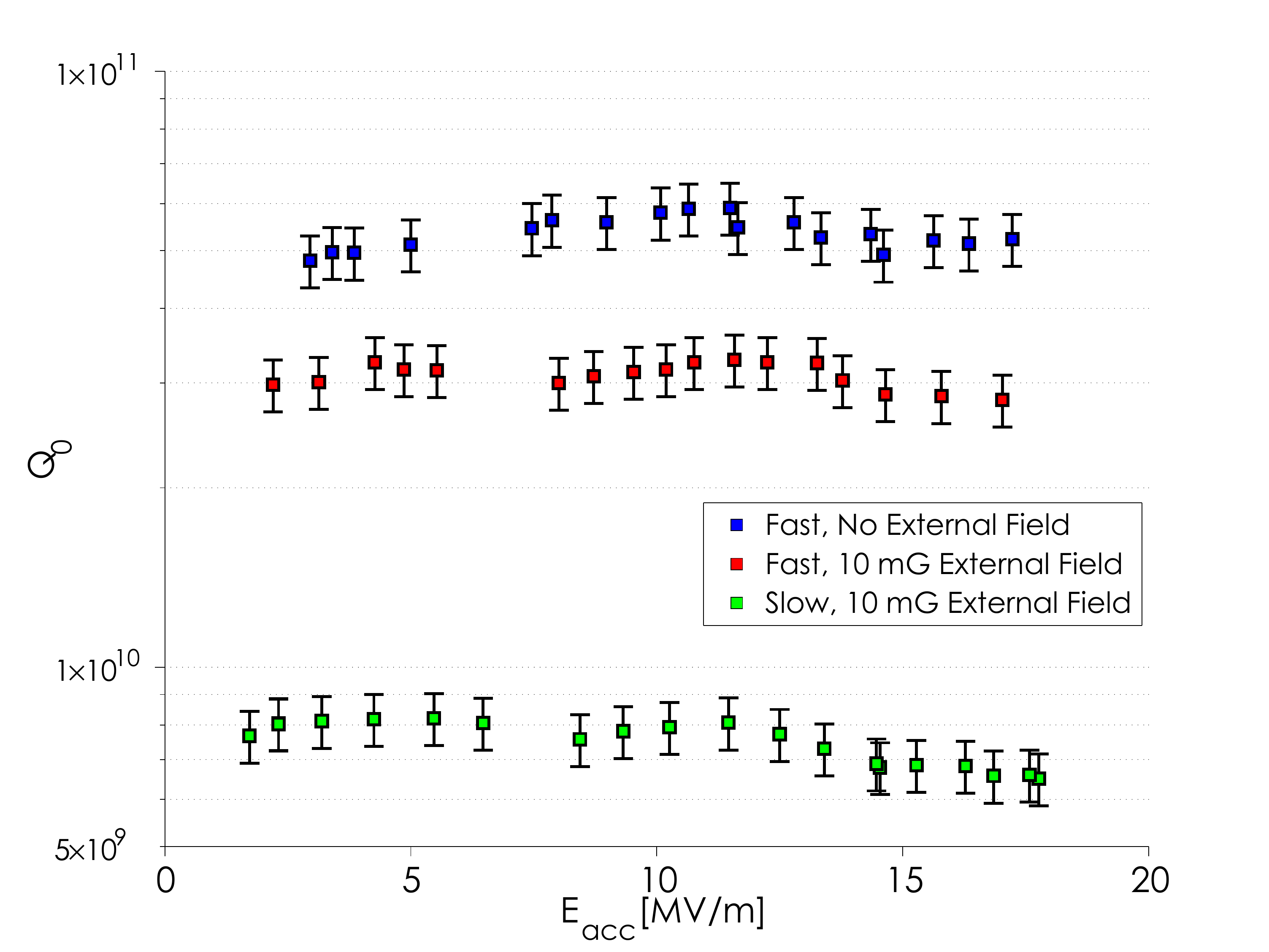}
\caption{The $Q_0$ vs $E_{acc}$ performance for HTC9-2 at 1.6~K for cool downs completed in $\sim$10 mG external magnetic field. The best fast cool down without applied solenoid field is also shown for reference. Uncertainty on  $E_{acc}$ is  10\%.} 
\label{fig18}
\end{figure}

As discussed before, HTC9-2 was assembled with a solenoid wrapped around the helium vessel of the cavity. The solenoid consisted of 10 equally spaced coils (5 Helmholtz coil pairs) in order to induce an approximately uniform external magnetic field parallel to the cavity axis. Using the coil, an average magnetic field of $\sim$10 mG was applied across the cavity and  the cavity was cooled twice, once fast (Fast 4), and once slow (Slow 1). The parameters of the cool downs are shown in Table \ref{tab3}. The purpose of these studies was to understand the sensitivity of nitrogen-doped cavities in cryomodule to ambient magnetic fields under realistic cool down conditions. Figure \ref{fig18} shows the 1.6 K $Q_0$ vs $E_{acc}$ for HTC9-2 during the second fast cool down (best cool down without applied solenoid field) for reference and the two cool downs with magnetic field (Fast 4, Slow 1). Under fast cool down ($\Delta T_{vertical}=7 \unit{K}$), residual resistance  increased by $4.6\pm0.9$~n$\Omega$ due  to the applied ambient magnetic field, as compared with the fast cool down that had similar vertical temperature gradients, but no additional magnetic field applied. With slow cool down in the same magnetic field, the increase was $36\pm7$~n$\Omega$ in residual resistance.

\begin{figure}
\centering
\includegraphics[scale=.27]{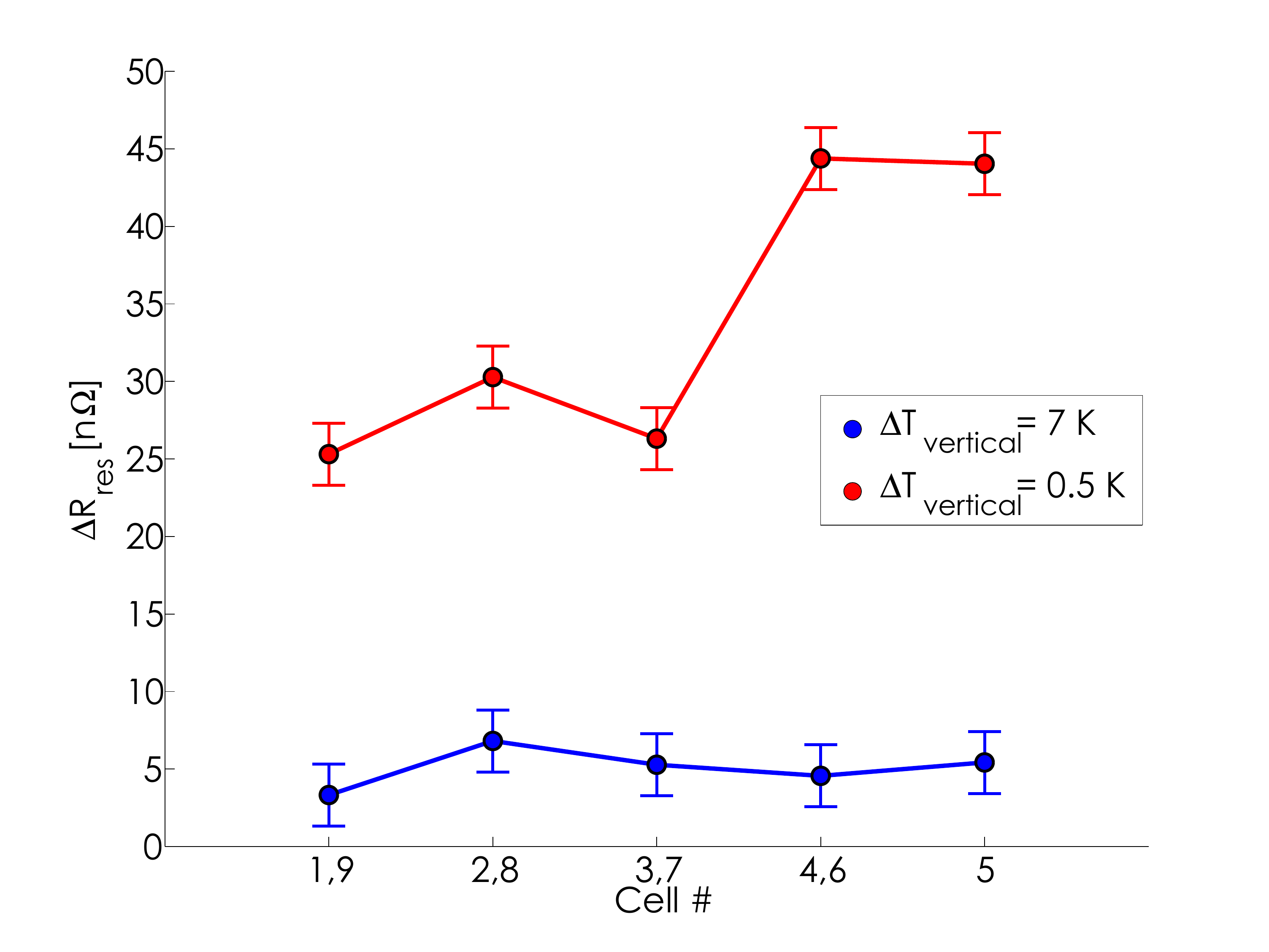}
\caption{The change in residual resistance per cell between the cool downs with 10 mG applied ambient magnetic field and the best fast cool down (Fast 2) without applied field.}
\label{fig19}
\end{figure}

By analyzing other TM010 mode data, in a manner similar to what was discussed above, the change in residual resistance per cell pair can be computed for the cool downs with applied ambient magnetic field. This change by cell is shown in Fig.~\ref{fig19}. From this data, we can make two conclusions: first that in slow cool down the largest change in residual resistance happened in the center cells of the cavity; and second, that large vertical temperature gradients result in significantly less flux trapping in all cells. 

Analyzing this data leads to the following sensitivities of nitrogen-doped cavities in a cryomodule to losses from trapped ambient magnetic fields: For fast cool down ($\Delta T_{vertical}=~7~\unit{K}$) the measured change in residual resistance for a given change in magnetic field is 0.5~n$\Omega$/mG. For slow cool down ($\Delta T_{vertical}=~0.5~\unit{K}$) the change in residual resistance for a given change in magnetic field is 3.8 n$\Omega$/mG. Assuming that during slow cool down there was near 100\% flux trapping, this value of 3.8~n$\Omega$/mG is consistent with flux trapping measurements on nitrogen-doped single-cell cavities at Cornell in which additional residual resistance was seen between 3 and 4~n$\Omega$ per mG of trapped flux \cite{DanLinacFlux}. These results allow for a first  prediction of the trapped flux contributions to residual resistance for the cavities in the LCLS-II cryomodules with fast cool down. Since the cryomodule specifications call for ambient magnetic fields at the cavity locations of no more that 5 mG, one can expect  an additional 2.5 n$\Omega$ of residual resistance with fast cool down  ($\Delta T_{vertical}=7$K). It is important to note that the measurements presented here were completed using an ILC helium tank. Further optimization in tank design may allow higher vertical spatial temperature gradients to be reached, which in turn would lower the sensitivity to ambient magnetic fields.

\section{Heater Study on HTC9-2}
Previous studies have demonstrated that the cool down dynamics around the critical temperature $T_c$ of niobium can have very significant impact on the average residual surface resistance of  dressed and undressed cavities. Measurements on a dressed cavity at Helmholtz Zentrum Berlin (HZB) showed that the residual resistance of the cavity  increased with larger longitudinal temperature gradients across the cavity near $T_c$ (based on the readings of  temperature sensors on the beam tubes outside of helium vessel) \cite{HZB1,HZB2}.  This effect was attributed to the additional magnetic field generated by thermoelectric currents flowing through the bimetal loop created by the cavity and titanium vessel, which gets trapped during the cool down through $T_c$. 
Due to the Seebeck effect, a spatial temperature gradient will drive a thermoelectric current,
\begin{equation}
\vec{J}=-\sigma(T)\left(\nabla V+S(T) \nabla T\right),
\label{eq1}
\end{equation}
where $S$ is the Seebeck coefficient, which depends on temperature and is different for different materials. Since  no magnetic field  probes were installed inside the LHe tank during the HZB test, no direct measurements  of the induced magnetic field by the thermoelectric currents were possible to support this explanation.  In apparent  contrast to the HZB results,  later vertical performance tests of dressed and undressed cavities showed the benefits of  fast cool down with large temperature gradients over the cavity \cite{AnnaCoolDown,DanLinacFlux,AlexMagField}. The measured reduction of residual resistance for fast cool downs was explained by a reduction in the trapped fraction of the ambient magnetic field by the larger temperature gradients at the normal- to superconducting transition. In addition, theoretical analysis showed  that the axial symmetry of the SRF cavities leads to very small thermoelectric induced magnetic fields in the relevant RF penetration layer at the inner cavity surface  \cite{Curtis}.  Therefore, in vertical cavity tests, these  thermoelectric induced magnetic fields usually do not cause a significant surface resistance degradation.

\begin{figure}
\centering
\includegraphics[scale=.27]{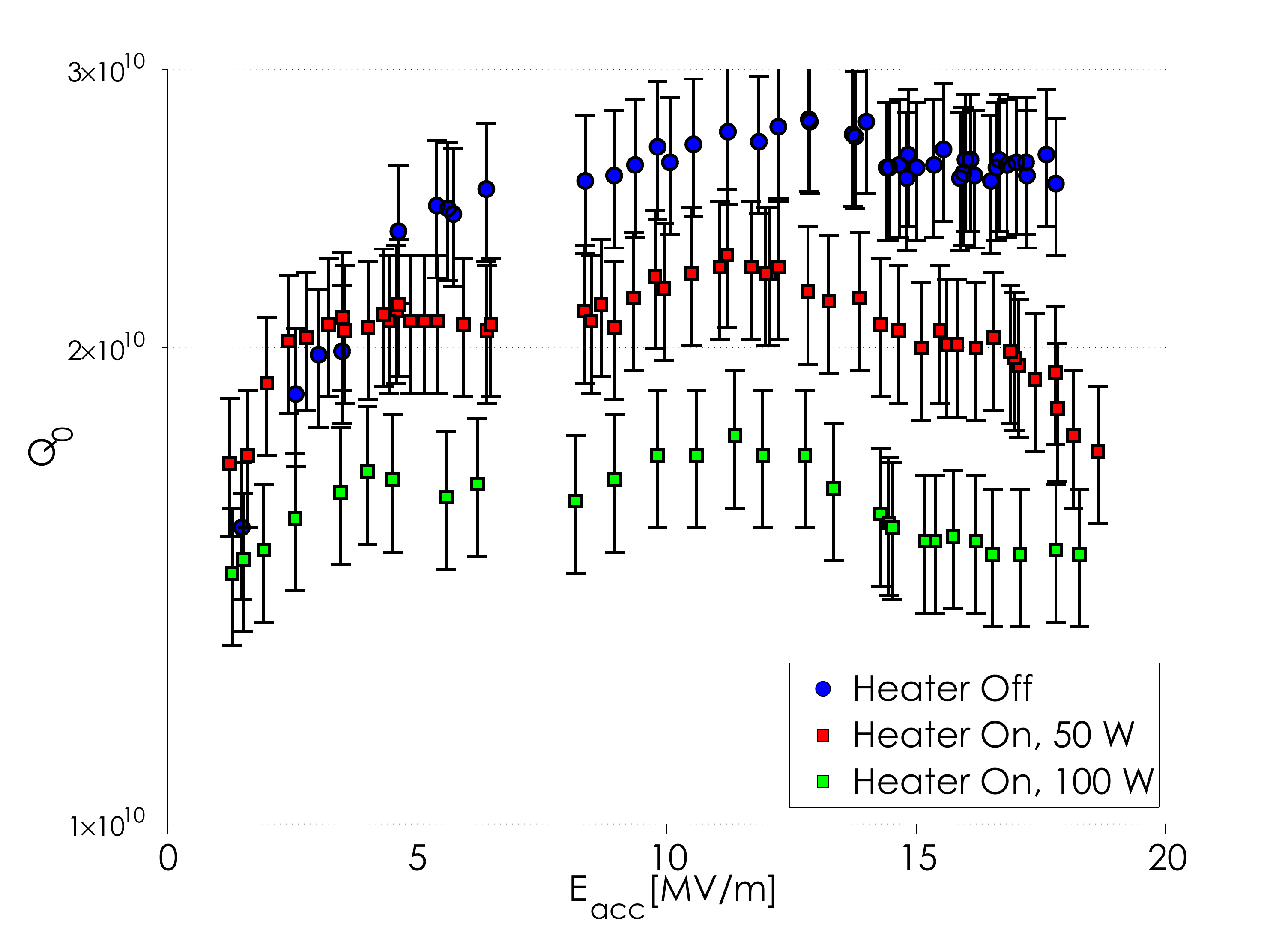}
\caption{The $Q_0$ vs $E_{acc}$ performance of HTC9-2 at 2.0 K. Uncertainty on $E_{acc}$ is 10\%.}
\label{fig21}
\end{figure}

To  study the trade-off in horizontal cavity performance between the potential detrimental effect  of  thermoelectric currents  generated by large thermal gradients and the benefits of large temperature gradients in reducing flux trapping, a heater was installed on the cell \#9 beam tube during the HTC9-2 test. The heater was then used to generate large longitudinal gradients during cool down, affecting both the magnitude of the thermoelectric current and thus the induced ambient magnetic field as well as the flux trapping. As can be seen in Table~\ref{tab3}, without heater the largest horizontal temperature gradient achieved was 9.4 K. With heater however, the horizontal gradient reached as high as $\sim$30 K. Clearly the heater was successful in inducing larger gradients. In total, two fast cool downs were performed with heater: one with the heater set to 50 W and one with it at 100 W. The resulting $Q_0$ vs $E_{acc}$ performances of the cavity at 2.0 K are shown in Fig.~\ref{fig21}. For reference the performance from the best cool down without heater is also shown. 

\begin{figure}
\centering
\includegraphics[width=80mm]{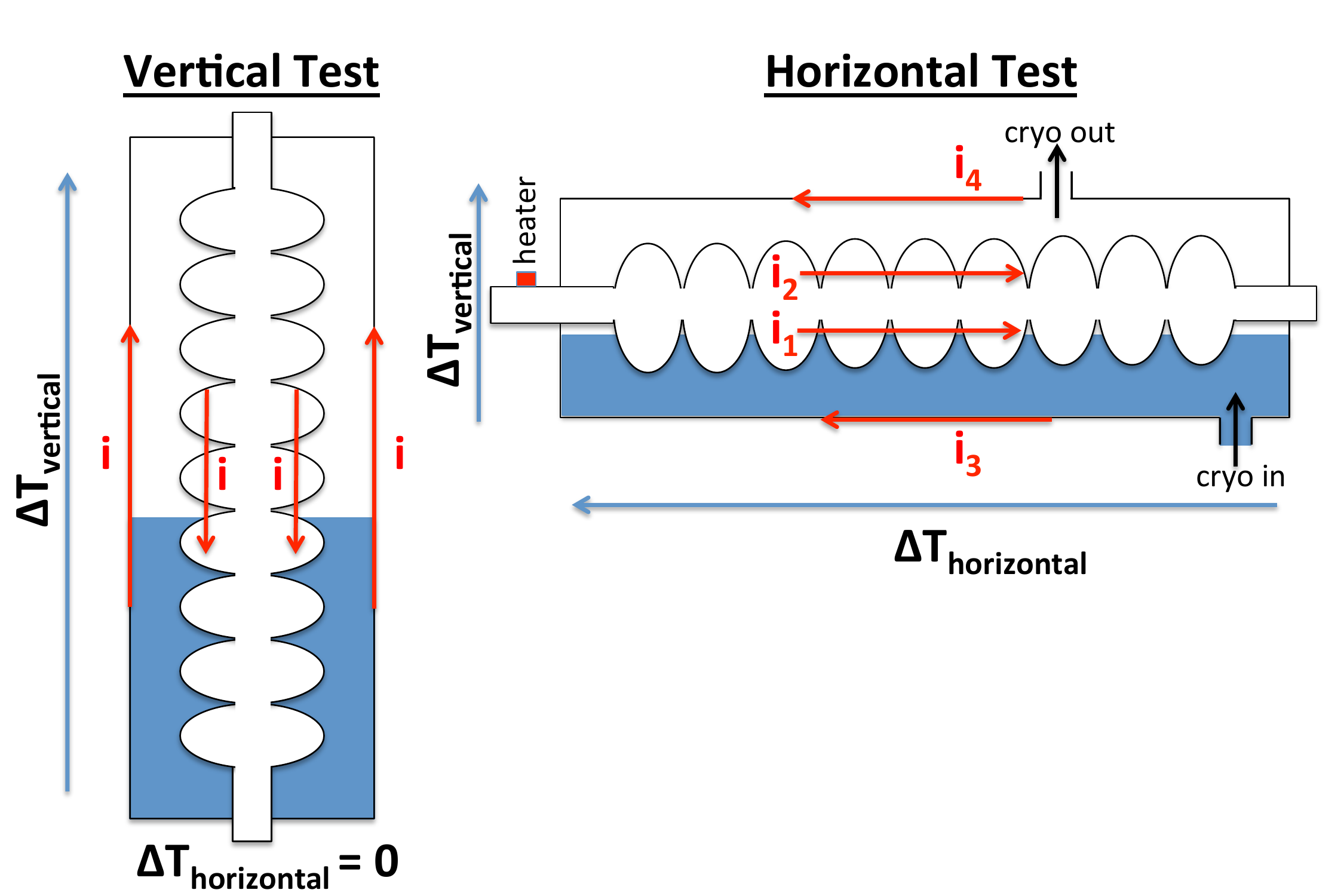}
\caption{Differences in thermoelectric currents in vertical and horizontal cavity tests. }
\label{fig21b}
\end{figure}

This experiment showed that the larger horizontal temperature gradients produced by the heater resulted in a significant degradation of the $Q_0$ even in fast cool downs. This can be understood by carefully analyzing the thermoelectric magnetic fields produced during cavity cool down. Larger horizontal temperature gradients will result in larger magnetic fields being present during cavity cool down. As shown in Table \ref{tab3}, the perpendicular magnetic field during the two cool downs in which the heater was used was as high as 62.8 mG at 10~K, compared with 0.3 mG during the cool down without heater that produced the best results. The fields generated by the currents in the heater itself and its current wires  were negligible. In vertical cavity tests, axial symmetry of the SRF cavities leads to very small thermoelectric induced magnetic fields in the  RF penetration layer, and thus to no significant increase in residual resistance due to trapped flux.  However, in horizontal cavity test this argument does not hold. For cavities placed horizontally with the cool down connections at the bottom of the liquid helium tanks, symmetry is broken since the cavities primarily cool from the bottom of the cells to the top of the cells; see  Fig.~\ref{fig7} and Fig.~\ref{fig8}. This leads to  vertical temperature gradients in addition to the horizontal temperature gradients. Since the electric conductivities of the titanium and the niobium strongly change with temperature, especially when the niobium becomes superconducting, the vertical temperature gradients result in non-symmetric thermoelectric currents through the cavity (higher currents at the bottom) and titanium tank as shown in Fig.~\ref{fig21b}.  These non-symmetric thermoelectric currents then can produce very significant magnetic fields in the RF penetration layer at the inner cavity wall, which partly get trapped and cause increased surface resistance. The HTC9-2 tests with heater show that the detrimental effect of the increased thermoelectric currents outweigh the reduction in the fraction of ambient field trapped  when the longitudinal temperature gradients are increased, thus overall resulting in lower $Q_0$. Therefore, in order to achieve low residual resistance and high $Q_0$,  horizontal cavities in cryomodules should be cooled with as small a horizontal temperature gradient as possible to reduce thermoelectric currents, while keeping  vertical temperature gradient as large as possible to reduce trapping of ambient magnetic fields.

\section{Conclusions}
Cornell has successfully completed the first horizontal tests of two nitrogen-doped cavities in a cryomodule. The HTC used for testing is very similar to the design of the LCLS-II cryomodule and thus serves as a powerful tool for studying how LCLS-II cavities will likely perform in the machine. We have achieved high $Q_0$, meeting and exceeding LCLS-II specifications. HTC9-1 achieved the highest 2.0 K $Q_0$ of 3.2\e{10} at 14~MV/m and HTC9-2 reached a $Q_0$ of 2.7\e{10} at 16 MV/m with a quench field of 20 MV/m, significantly higher than the LCLS-II specification. From these tests, important lessons were learned about the performance of nitrogen-doped cavities when assembled horizontally.

Both cryomodule tests showed a small degradation in $Q_0$ between vertical and horizontal test. This change comes from an increase in residual resistance of 1 to 3~n$\Omega$,  likely resulting from higher flux trapping in horizontal test than is achieved in vertical test, even with fast cool down. These studies provide the first data on the degradation margin for $Q_0$ in nitrogen-doped cavities from vertical to horizontal test.

A solenoid was used to apply an external longitudinal magnetic field during HTC9-2. Two cool downs (one fast, one slow) were completed in an average magnetic field of 10 mG. From these studies, the sensitivity of cavity residual resistance to ambient magnetic field was measured. It was found that with the Cornell cool down procedure (fast, $\Delta T_{vertical}=7\unit{K}$), there was an increase of 0.5 n$\Omega$ of residual resistance per additional mG of ambient magnetic field. With slow cool down, the increase was 3.8 n$\Omega$/mG, consistent with flux trapping measurements on single-cell cavities. These measurements allow a prediction for the cavities in the LCLS-II cryomodule: an ambient magnetic field of 5 mG is expected to result in an additional 2.5 n$\Omega$ of residual resistance from trapped flux using the best fast cool down procedure achieved so far. This value is well within the $\sim$5~n$\Omega$ residual resistance budget for the LCLS-II cavities, assuming a typical 2.0~K BCS surface resistance at 16~MV/m of $\sim$5~n$\Omega$ for nitrogen doped cavities. Further optimization of the LCLS-II doping recipe may result in small changes to the sensitives to ambient magnetic field reported here.

A heater was placed on the cell \#9 beam tube during HTC9-2 in order to induce large horizontal temperature gradients. It was found that the large horizontal temperature gradients resulted in very large magnetic fields from thermoelectric currents that negatively impacted the $Q_0$ of the cavity, even with large vertical temperature gradients during cool down in these tests.  These studies suggest that in order to achieve low residual resistance and high $Q_0$ in a cryomodule, the cavities should be cooled with a large vertical temperature gradient (for reduction of flux pinning) and a small horizontal temperature gradient (to minimize fields from thermal currents). The LCLS-II helium tank should be optimized for this type of cooling with the liquid helium inlets being placed symmetrically on both ends of the cavity.

These measurements represent the first tests of nitrogen-doped cavities in a cryomodule. Together with the ongoing single and 9-cell high $Q_0$ work at Cornell, Fermilab, and Jefferson Lab, they show a strong support for the viability of nitrogen-doped cavities to be used in the LCLS-II project. Future work will continue on studying thermoelectric current and cool down effects on cavities and a third HTC test with a 9-cell cavity with the LCLS-II helium tank will be completed in the near future.

\section{ACKNOWLEDGMENTS}
The authors would like to thank the entire FNAL crew for preparing and dressing the cavity before shipping to Cornell. We would also like to thank the technicians at CLASSE for preparing the HTC for test in record time and the Cornell cryogenics group for staying late many nights in order to conduct thermal cycles.

\bibliographystyle{unsrt}
\bibliography{htclclsbib}

\end{document}